\documentclass[aps,prd,twocolumn,tightenlines,superscriptaddress,showpacs,amsmath,amssymb]{revtex4}
\usepackage{graphicx} %
\usepackage{epstopdf} %
\usepackage{dcolumn} %
\usepackage{xcolor} %
\usepackage{amsmath,amssymb} %
\usepackage{hyperref} %
\usepackage[utf8]{inputenc} %
\usepackage[english]{babel} %
\usepackage[mathlines]{lineno} %
\usepackage{blindtext} %
\usepackage{ifthen} %
\usepackage{orcidlink} %

\usepackage[noabbrev,capitalise]{cleveref} %

\usepackage{afterpage} %
\usepackage{rotating,booktabs,multirow} %

\usepackage{overpic}

\graphicspath{{figures/}} %

\newboolean{articletitles}
\setboolean{articletitles}{true} %

\newcommand*\patchAmsMathEnvironmentForLineno[1]{%
\expandafter\let\csname old#1\expandafter\endcsname\csname #1\endcsname
\expandafter\let\csname oldend#1\expandafter\endcsname\csname
end#1\endcsname
 \renewenvironment{#1}%
   {\linenomath\csname old#1\endcsname}%
   {\csname oldend#1\endcsname\endlinenomath}%
}
\newcommand*\patchBothAmsMathEnvironmentsForLineno[1]{%
  \patchAmsMathEnvironmentForLineno{#1}%
  \patchAmsMathEnvironmentForLineno{#1*}%
}
\AtBeginDocument{%
\patchBothAmsMathEnvironmentsForLineno{equation}%
\patchBothAmsMathEnvironmentsForLineno{align}%
\patchBothAmsMathEnvironmentsForLineno{flalign}%
\patchBothAmsMathEnvironmentsForLineno{alignat}%
\patchBothAmsMathEnvironmentsForLineno{gather}%
\patchBothAmsMathEnvironmentsForLineno{multline}%
\patchBothAmsMathEnvironmentsForLineno{eqnarray}%
}

\RequirePackage{xspace}
\RequirePackage{relsize}

\def\belletwo{\mbox{Belle~II}\xspace}

\def\babar{\mbox{\slshape B\kern-0.1em{\smaller A}\kern-0.1em
    B\kern-0.1em{\smaller A\kern-0.2em R}}\xspace}

\def\Ppi         {\ensuremath{\pi}\xspace}

\mathchardef\PDelta="7101
\mathchardef\PXi="7104
\mathchardef\PLambda="7103
\mathchardef\PSigma="7106
\mathchardef\POmega="710A
\mathchardef\PUpsilon="7107
                 
\def\PB      {\ensuremath{B}\xspace}                 
                 
\def\PD      {\ensuremath{D}\xspace}

\def\PK      {\ensuremath{K}\xspace}

\def\Pc      {\ensuremath{c}\xspace}                 
                 
\def\Pe      {\ensuremath{e}\xspace}

\def\Pi      {\ensuremath{i}\xspace}

\def\Pq      {\ensuremath{q}\xspace}

\def\epem       {\ensuremath{\Pe^+\Pe^-}\xspace}

\def\quark     {\ensuremath{\Pq}\xspace}
\def\quarkbar  {\ensuremath{\overline \quark}\xspace}
\def\qqbar     {\ensuremath{\quark\quarkbar}\xspace}

\def\cquark    {\ensuremath{\Pc}\xspace}
\def\cquarkbar {\ensuremath{\overline \cquark}\xspace}
\def\ccbar     {\ensuremath{\cquark\cquarkbar}\xspace}

\def\pion  {\ensuremath{\Ppi}\xspace}
\def\piz   {\ensuremath{\pion^0}\xspace}

\def\pip   {\ensuremath{\pion^+}\xspace}
\def\pim   {\ensuremath{\pion^-}\xspace}

\def\kaon  {\ensuremath{\PK}\xspace}
\def\Kbar  {\kern 0.2em\overline{\kern -0.2em \PK}{}\xspace}%

\def\Kz    {\ensuremath{\kaon^0}\xspace}
\def\Kzb   {\ensuremath{\Kbar^0}\xspace}
\def\KzKzb {\ensuremath{\Kz \kern -0.16em \Kzb}\xspace}
\def\Kp    {\ensuremath{\kaon^+}\xspace}
\def\Km    {\ensuremath{\kaon^-}\xspace}

\def\KpKm  {\ensuremath{\Kp \kern -0.16em \Km}\xspace}
\def\KS    {\ensuremath{\kaon^0_{\rm\scriptscriptstyle S}}\xspace} 
\def\KL    {\ensuremath{\kaon^0_{\rm\scriptscriptstyle L}}\xspace}

\def\D       {\ensuremath{\PD}\xspace}
\def\Dbar    {\kern 0.2em\overline{\kern -0.2em \PD}{}\xspace}%

\def\Dz      {\ensuremath{\D^0}\xspace}
\def\Dzb     {\ensuremath{\Dbar^0}\xspace}
\def\DzDzb   {\ensuremath{\Dz {\kern -0.16em \Dzb}}\xspace}
\def\Dp      {\ensuremath{\D^+}\xspace}
\def\Dm      {\ensuremath{\D^-}\xspace}

\def\DpDm    {\ensuremath{\Dp {\kern -0.16em \Dm}}\xspace}
\def\Dstar   {\ensuremath{\D^*}\xspace}

\def\Dstarp  {\ensuremath{\D^{*+}}\xspace}
\def\Dstarm  {\ensuremath{\D^{*-}}\xspace}

\def\B       {\ensuremath{\PB}\xspace}
\def\Bbar    {\ensuremath{\kern 0.18em\overline{\kern -0.18em \PB}{}}\xspace}%

\def\Y#1S{\ensuremath{\PUpsilon{(#1S)}}\xspace}%

\def\Lbar {\ensuremath{\kern 0.1em\overline{\kern -0.1em\PLambda}}\xspace}

\def\to                 {\ensuremath{\rightarrow}\xspace}

\def\CP                {\ensuremath{C\!P}\xspace}

\newcommand{\tev}{\ensuremath{\mathrm{\,Te\kern -0.1em V}}\xspace}
\newcommand{\gev}{\ensuremath{\mathrm{\,Ge\kern -0.1em V}}\xspace}
\newcommand{\mev}{\ensuremath{\mathrm{\,Me\kern -0.1em V}}\xspace}
\newcommand{\kev}{\ensuremath{\mathrm{\,ke\kern -0.1em V}}\xspace}
\newcommand{\ev}{\ensuremath{\mathrm{\,e\kern -0.1em V}}\xspace}
\newcommand{\gevc}{\ensuremath{{\mathrm{\,Ge\kern -0.1em V\!/}c}}\xspace}
\newcommand{\mevc}{\ensuremath{{\mathrm{\,Me\kern -0.1em V\!/}c}}\xspace}
\newcommand{\gevcc}{\ensuremath{{\mathrm{\,Ge\kern -0.1em V\!/}c^2}}\xspace}
\newcommand{\gevgevcccc}{\ensuremath{{\mathrm{\,Ge\kern -0.1em V^2\!/}c^4}}\xspace}
\newcommand{\mevcc}{\ensuremath{{\mathrm{\,Me\kern -0.1em V\!/}c^2}}\xspace}

\def\cm   {\ensuremath{\rm \,cm}\xspace}

\def\invfb   {\ensuremath{\mbox{\,fb}^{-1}}\xspace}

\newcommand{\chisq}{\ensuremath{\chi^2}\xspace}

\def\gsim{{~\raise.15em\hbox{$>$}\kern-.85em
          \lower.35em\hbox{$\sim$}~}\xspace}
\def\lsim{{~\raise.15em\hbox{$<$}\kern-.85em
          \lower.35em\hbox{$\sim$}~}\xspace}

\def\sPlot{\mbox{\em sPlot}}

\newcommand{\vs}{\mbox{\itshape vs.}\xspace}
 
\newcommand{\Acp}{\ensuremath{A_{\CP}}\xspace}

\newcommand{\pis}{\ensuremath{\pi_{\rm s}}\xspace}

\makeatletter%
\begin{document}
\includegraphics[width=3cm]{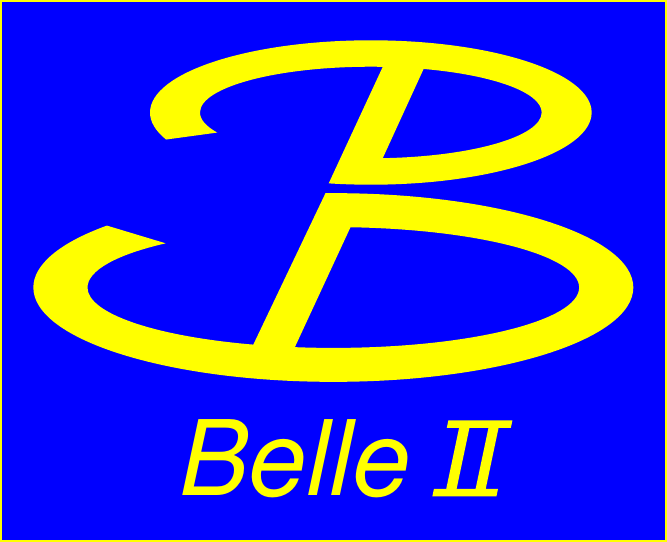}\vspace*{-1.9cm}

\begin{flushright}
Belle II Preprint 2025-009\\
KEK Preprint 2025-7
\end{flushright}\vspace{1.5cm}

\title{%
%
{\LARGE\bfseries\boldmath Measurement of the time-integrated \CP asymmetry in $\Dz\to\piz\piz$ decays at Belle II}
 %
}
%
%
%
%
%
%
  \author{I.~Adachi\,\orcidlink{0000-0003-2287-0173}} %
  \author{Y.~Ahn\,\orcidlink{0000-0001-6820-0576}} %
  \author{N.~Akopov\,\orcidlink{0000-0002-4425-2096}} %
  \author{S.~Alghamdi\,\orcidlink{0000-0001-7609-112X}} %
  \author{M.~Alhakami\,\orcidlink{0000-0002-2234-8628}} %
  \author{A.~Aloisio\,\orcidlink{0000-0002-3883-6693}} %
  \author{N.~Althubiti\,\orcidlink{0000-0003-1513-0409}} %
  \author{K.~Amos\,\orcidlink{0000-0003-1757-5620}} %
  \author{M.~Angelsmark\,\orcidlink{0000-0003-4745-1020}} %
  \author{N.~Anh~Ky\,\orcidlink{0000-0003-0471-197X}} %
% ...
%
\collaboration{The Belle II Collaboration}
 \begin{abstract}
%
%
\noindent We measure the time-integrated \CP asymmetry, \Acp, in $\Dz\to\piz\piz$ decays reconstructed in $\epem\to \ccbar$ events collected by \belletwo during 2019--2022. The data corresponds to an integrated luminosity of 428\invfb. The \Dz decays are required to originate from the flavor-conserving $\Dstarp \to \Dz \pip$ decay to determine the charm flavor at production time. Control samples of $\Dz\to \Km \pip$ decays, with or without an associated pion from a \Dstarp decay, are used to correct for detection asymmetries. The result, $\Acp(\Dz\to\piz\piz) = (0.30\pm 0.72\pm 0.20)\%$, where the first uncertainty is statistical and the second systematic, is consistent with \CP symmetry. 
 \end{abstract}

\maketitle

%
%
\section{Introduction}
Searches for charge-parity (\CP) violation in charm decays offer the opportunity to probe for physics beyond the standard model. Since, to a good approximation, the third generation of quarks can be ignored when describing charm transitions, \CP-violation effects are expected to be tiny, with standard-model asymmetries typically ranging from $10^{-3}$ to $10^{-4}$~\cite{Golden:1989qx,Buccella:1994nf,Bianco:2003vb,Grossman:2006jg,Artuso:2008vf}. New dynamics may alter the expected decay or, for neutral mesons, flavor-mixing rates and introduce larger \CP-violation effects. After decades of experimental efforts, \CP violation in charm transitions has been observed only as the difference between the time-integrated \CP asymmetries of $D^0\to\Kp\Km$ and $D^0\to\pip\pim$ decays~\cite{Aaij:2019kcg}, with strong evidence that \CP violation occurs mainly in the direct decay $\Dz\to\pip\pim$~\cite{LHCb:2022lry}. (Throughout this paper, charge-conjugate modes are implied unless stated otherwise.) These results can be interpreted either as a sign of new dynamics or as an enhancement of the subleading amplitude mediated by non-perturbative QCD~\cite{Chala:2019,Dery:2019ysp,Calibbi:2019bay,Grossman:2019,Cheng:2019ggx,Buras:2021rdg,Schacht:2021jaz,Bediaga:2022sxw,Pich:2023kim,Gavrilova:2023fzy,Lenz:2023rlq}. Moreover, they indicate a larger than expected breaking of the $U$-spin symmetry, which requires the \CP asymmetries of the direct $\Dz\to\Kp\Km$ and $\Dz\to\pip\pim$ decays to have equal magnitudes and opposite signs: $|A_{\CP}^\text{dir}(\Dz\to\Kp\Km)|=-|A_{\CP}^\text{dir}(\Dz\to\pip\pim)|$~\cite{Schacht:2022}.

Precise measurements of \CP asymmetries in the isospin-related $D^+ \to \pi^+\pi^0$ and $D^0 \to \pi^0\pi^0$ modes can help to clarify the picture~\cite{Grossman:2012,Bevan:2013xla}. In the standard model, direct \CP violation in Cabibbo-suppressed charm decays is generated by the interference of a leading tree-level amplitude and a suppressed $\Delta I=1/2$ QCD-penguin amplitude, which changes isospin by half a unit. The $\pip\piz$ final state has isospin $I=2$ and cannot be reached from the $I=1/2$ initial state via a $\Delta I=1/2$ transition. Hence, in the standard model, $\Acp^\text{dir}(D^+\to\pi^+\piz)=0$. On the contrary, the $\pi^0\pi^0$ and $\pi^+\pi^-$ final states can have $I=0$ or $I=2$ and hence can have nonzero direct \CP asymmetries in the standard model. As an example, recent experimental results imply that $\Acp^\text{dir}(D^0\to\piz\piz)$ may be as large as $1\%$~\cite{Wang:2022}. The following sum-rule helps determine the source of \CP violation:~\cite{Grossman:2012,HFLAV}
\begin{multline}\label{eq:sum-rule}
    R = \frac{A_{\CP}^\text{dir}(D^0\to\pi^+\pi^-)}{1+\frac{\tau_{D^0}}{\mathcal{B}_{+-}} \left( \frac{\mathcal{B}_{00}}{\tau_{D^0}} - \frac{2}{3} \frac{\mathcal{B}_{+0}}{\tau_{D^+} } \right)} + \frac{A_{\CP}^\text{dir}(D^0\to\pi^0\pi^0)}{1+\frac{\tau_{D^0}}{\mathcal{B}_{00}} \left( \frac{\mathcal{B}_{+-}}{\tau_{D^0}} - \frac{2}{3} \frac{\mathcal{B}_{+0}}{\tau_{D^+} } \right)} \\ + \frac{A_{\CP}^\text{dir}(D^+\to\pi^+\pi^0)}{1- \frac{3}{2} \frac{\tau_{D^+}}{\mathcal{B}_{+0}} \left( \frac{\mathcal{B}_{00}}{\tau_{D^0}} + \frac{\mathcal{B}_{+-}}{\tau_{D^0} } \right)}\,,
\end{multline}
where $\tau_{D^0}$ and $\tau_{D^+}$ are the lifetimes of $D^0$ and $D^+$ mesons, and, $\mathcal{B}_{+-}$, $\mathcal{B}_{00}$ and $\mathcal{B}_{+0}$ are the \D meson branching fractions to $\pi^+\pi^-$, $\pi^0\pi^0$ and $\pi^+\pi^0$ decays. If $R$ is measured to be nonzero, then \CP violation arises from the $\Delta I=1/2$ amplitude. If $R$ is zero, and at least one of the direct \CP asymmetries is nonzero, then \CP violation arises from a beyond-standard-model $\Delta I=3/2$ amplitude. The value is measured to be $R=(0.9\pm 3.1)\times10^{-3}$, where the uncertainty is dominated by the asymmetry of the $\piz\piz$ final state~\cite{HFLAV}. The most precise measurement of the time-integrated \CP asymmetry in $D^0\to\piz\piz$ available to date comes from the Belle collaboration, $(-0.03\pm0.64 \pm0.10)\%$~\cite{Nisar:2014}, where the first uncertainty is statistical and the second systematic. (Indirect \CP violation induced by \Dz-\Dzb mixing is precisely measured in other channels~\cite{LHCb:2021vmn} and subtracted from the time-integrated asymmetry when computing $R$~\cite{Grossman:2006jg,CDF:2011ejf}.) Belle~II is the only existing experiment capable of improving this measurement and the determination of the sum rule.

In this paper, we report a measurement of the time-integrated \CP asymmetry in $\Dz\to\piz\piz$ decays,
\begin{equation}
\Acp(\Dz\to\piz\piz) = \frac{\Gamma(\Dz\to\piz\piz)-\Gamma(\Dzb\to\piz\piz)}{\Gamma(\Dz\to\piz\piz)+\Gamma(\Dzb\to\piz\piz)}
\end{equation}
where $\Gamma$ is the decay-time integrated rate, which includes effects due to \Dz-\Dzb mixing. The measurement uses $\epem\to\ccbar$ data collected by Belle~II between 2019 and 2022, which correspond to an integrated luminosity of 428\invfb. To determine the production flavor of the neutral \D meson, we reconstruct $\Dz\to\piz\piz$ decays originating from $\Dstarp \to \Dz \pi^+$ decays. We refer to \Dz
mesons originating from identified \Dstarp decays as tagged decays, and to the low-momentum flavor-tagging pion as the ``soft'' pion or \pis. To measure the \CP asymmetry, we determine the observed asymmetry between the number of detected decays of opposite flavor,
\begin{multline}
A^{\piz\piz} = \\ \resizebox{0.45\textwidth}{!}{$\frac{N[\Dstarp\to(\Dz\to\piz\piz)\pip]-N[\Dstarm\to(\Dzb\to\piz\piz)\pim]}{N[\Dstarp\to(\Dz\to\piz\piz)\pip]+N[\Dstarm\to(\Dzb\to\piz\piz)\pim]}\,.
$}
\end{multline}
This raw asymmetry is a sum of contributions from \CP violation, \Acp; from the forward-backward asymmetric production of \Dstarp mesons in $\epem\to\ccbar$ events, $A_{\rm P}^{\Dstarp}$; and from charge asymmetries in the detection efficiency of positive and negative soft pions, $A_\epsilon^{\pis}$:
\begin{equation}
A^{\piz\piz} = \Acp(\Dz\to\piz\piz)+A_{\rm P}^{\Dstarp}+A_{\epsilon}^{\pis}\,.    
\end{equation}
We use high-yield control samples of tagged and untagged Cabibbo-favored $\Dz\to\Km\pip$ decays, where \CP violation can be neglected, to correct for the instrumental asymmetries. The raw asymmetries of these decays are
\begin{align}
A^{K\pi} = A_{\rm P}^{\Dstarp}+A_{\epsilon}^{\pis}+A_{\epsilon}^{K\pi}
\intertext{and}
A^{K\pi,\mathrm{untag}} = A_{\rm P}^{\Dz}+A_{\epsilon}^{K\pi}\,,
\end{align}
where $A_{\epsilon}^{K\pi}$ is the detection asymmetry of the $\Km\pip$ pair. In $\epem\to\ccbar$ events, charmed mesons are produced with a forward-backward asymmetry due to $\gamma$-$Z^{0}$ interference and higher-order effects~\cite{Berends:1973fd,Brown:1973ji,Cashmore:1985vp}. Since the acceptance of the Belle II detector is not the same in the forward and backward directions, a charge asymmetry in the production of charmed mesons remains. Being an odd function of the cosine of the charmed-meson polar angle in the \epem center-of-mass system, $\cos\theta_{\rm cms}$, this production asymmetry is suppressed by averaging the raw asymmetries from forward and backward decays:
\begin{equation}\label{eq:aprime}
A^{\prime\,f} = \frac{A^f(\cos\theta_{\rm cms}<0)+A^f(\cos\theta_{\rm cms}>0)}{2}\,,
\end{equation}
for $f=\piz\piz$; $K\pi$; $K\pi$, untag. Assuming that the detection asymmetries for different decays can be made identical by weighting the events, the \CP asymmetry in $\Dz\to\piz\piz$ is then
\begin{equation} \label{eq:diff_acp}
\Acp(\Dz\to\piz\piz) = A^{\prime\,\piz\piz} - A^{\prime\,K\pi} + A^{\prime\,K\pi,\mathrm{untag}}\,,
\end{equation}
where the difference between the first two terms cancels the soft-pion detection asymmetry, and the difference between the second and third terms cancels the $\Km\pip$ detection asymmetry. To avoid potential bias, the measured value of $A^{\piz\piz}$ remained unexamined until the entire analysis procedure was finalized and all uncertainties were determined.

The paper is organized as follows. \Cref{sec:detector} provides an overview of the Belle~II detector and of the simulation samples used in the measurement. The reconstruction and selection of both the signal $\Dz\to\piz\piz$ and control $\Dz\to\Km\pip$ decays are presented in \cref{sec:selection}. \Cref{sec:equalization} discusses the weighting of the control modes to match the kinematic distributions of the signal mode and ensure an accurate cancellation of the detector asymmetries. Determination of the raw asymmetries is covered in \cref{sec:fit}, followed by a discussion of the systematic uncertainties affecting the measurement in \cref{sec:systematics}. Final results are presented in \cref{sec:results}, followed by concluding remarks.

\section{Belle~II detector and simulation\label{sec:detector}}
The Belle~II detector~\cite{b2tech,Kou:2018nap} operates at the SuperKEKB asymmetric-energy $\epem$ collider~\cite{Akai:2018mbz}. It has a cylindrical geometry and consists of a silicon vertex detector comprising two inner layers of pixel detectors and four outer layers of double-sided strip detectors, a 56-layer central drift chamber, a time-of-propagation detector, an aerogel ring-imaging Cherenkov detector, and an electromagnetic calorimeter made of CsI(Tl) crystals, all located inside a 1.5 T superconducting solenoid. A flux return outside the solenoid is instrumented with resistive-plate chambers and plastic scintillator modules to detect muons and \KL mesons. For the data used in this paper only part of the second layer of the pixel detector, covering 15\% of the azimuthal angle, was installed. The $z$ axis of the laboratory frame is defined as the central axis of the solenoid, with its positive direction determined by the direction of the electron beam.

We use simulated event samples to identify sources of background, optimize selection criteria, match the kinematic distributions of signal and control decays, determine fit models, and validate the analysis procedure. We generate $\epem\to\Upsilon(4S)$ events and simulate particle decays with the \textsc{EvtGen} generator~\cite{Lange:2001uf}; we generate continuum $\epem\to\qqbar$ (where $q$ is a $u$, $d$, $c$, or $s$ quark) with \textsc{KKMC}~\cite{Jadach:1999vf} and \textsc{Pythia8}~\cite{Sjostrand:2014zea}; we simulate final-state radiation with \textsc{Photos}~\cite{Barberio:1990ms,Barberio:1993qi}; we simulate detector response using \textsc{Geant4}~\cite{Agostinelli:2002hh}. Beam backgrounds are taken into account by overlaying random-trigger data. The data and simulation samples are processed using the Belle II analysis software framework~\cite{Kuhr:2018lps,basf2-zenodo}.

\section{Event selection\label{sec:selection}}

\subsection{$\Dz\to \piz\piz$ signal sample}
We reconstruct photon candidates from localized energy deposits, clusters, in the electromagnetic calorimeter that are consistent with an electromagnetic shower based on pulse-shape discrimination~\cite{Longo:2020zqt}. The clusters should have a polar angle in the laboratory frame, $\theta$, within the acceptance of the drift chamber ($17<\theta<150^\circ$) to ensure that they are not matched to tracks. They must contain energy from at least two crystals and an energy deposit greater than 25\mev if located in the forward ($12.4<\theta<31.4^\circ$) or barrel ($32.2<\theta<128.7^\circ$) regions, and greater than 40\mev if in the backward region ($130.7<\theta<155.7^\circ$). Two photon candidates are then combined to form a neutral pion candidate if their mass is in $[105,150]\mevcc$ (the typical diphoton mass resolution is 7\mevcc). Pairs of \piz candidates are combined to form $\Dz\to\piz\piz$ candidates and the dipion mass, $m(\piz\piz)$, is required to be in [1.6,2.1]\gevcc. 

Soft pion candidates are charged particles that are in the acceptance of the drift chamber and originate from the \epem interaction region, meaning that they have longitudinal and transverse distances of closest approach to the \epem interaction point smaller than 3\cm and 1\cm, respectively. A \Dz candidate and a soft pion candidate are then combined to form $\Dstarp\to\Dz\pip$ candidates. The difference between the reconstructed \Dstarp and \Dz masses, $\Delta m$, must not exceed 160\mevcc. The \Dstarp candidates are subject to a fit that uses both kinematic and spatial information and constrains the \Dstarp vertex to the measured position of the beam interaction point~\cite{Krohn:2019dlq}. Only candidates with successful fits and having \chisq probabilities larger than 0.001 are retained for subsequent analysis. The momentum of the \Dz candidate in the \epem center-of-mass system is required to exceed 2.5\gevc to suppress events where the \Dz candidate originates from the decay of a \B meson, since \Dz mesons from this source may have a different production asymmetry.

We use a histogram-based gradient-boosting decision tree~\cite{scikit-learn,histogram-gbdt} (HBDT) to discriminate signal from background made of fake \piz candidates and neutral pions originating from unrelated processes. The HBDT uses 14 input variables: the momentum of the \Dz candidate in the \epem center-of-mass system, the momenta of the two \piz candidates, the transverse momentum of the soft pion, the opening angle between the two photons of each \piz decay, and for every photon in the reconstruction chain, the output of two multivariate classifiers that use shower-related variables to suppress energy clusters due to beam-related background or hadronic showers. The HBDT is trained and tested on independent samples of simulated decays to mitigate overtraining. To suppress background, the HBDT output is required to be larger than a threshold value that maximizes the figure-of-merit $N_{\mathrm{sig}}/\sqrt{N_{\mathrm{sig}}+N_{\mathrm{bkg}}}$, where $N_{\mathrm{sig}}$ and $N_{\mathrm{bkg}}$ are signal and background yields in the region defined by $m(\piz\piz)\in [1.78,1.92]\gevcc$ and $\Delta m\in [0.144,0.147]\gevcc$, as estimated using simulation. The HBDT requirement rejects 95\% of the background while retaining 85\% of the signal. After this selection, 9\% of events have multiple \Dstarp candidates. In such cases, we retain the candidate with the highest HBDT output value. According to simulation, this criterion selects the correct candidate in 84\% of cases. 

\subsection{$\Dz\to\Km\pip$ control samples}
The reconstruction of the control mode starts by selecting events that are inconsistent with Bhabha scattering and have at least three tracks with transverse momentum larger than $0.2\gevc$ that originate from the \epem interaction region.

Charged kaon and pion candidates are required to be in the acceptance of the drift chamber, to originate from the \epem interaction region, and to be identified as kaons and pions, respectively. Particle identification is performed using information from all subdetector systems, except for the pixel detector~\cite{PID}. The kaon and pion-identification efficiencies are above 83\% and 93\%, respectively, with corresponding pion-as-kaon and kaon-as-pion misidentification rates below 5\% and 10\%. A negatively charged kaon candidate and a positively charged pion candidate are combined to form $\Dz\to\Km\pip$ candidates with an invariant mass, $m(\Km\pip)$, in $[1.814,1.914]\gevcc$. A vertex-kinematic fit~\cite{Krohn:2019dlq} requiring the \Dz candidate momentum to point back to the \epem interaction region is performed, and only candidates with fit $\chi^2$ probabilities in excess of 0.001 are retained. To suppress events where the \Dz candidate originates from the decay of a \B meson, the momentum of the \Dz candidate in the \epem center-of-mass system is required to exceed 2.5\gevc. All candidates are retained in the $<1\%$ of events where more than one \Dz candidate is reconstructed. These \Dz candidates form the untagged control sample. 

The tagged control sample is obtained by combining \Dz and soft-pion candidates to form $\Dstarp \to \Dz \pi_s^+$ candidate decays with $\Delta m$ values in the range $[142.0,148.8]\mevcc$. The soft pions are selected with the same requirements as for the signal mode. We suppress $\Dstarp\to\Dz\pi_s^+$ decays with a mis- or partially reconstructed \Dz candidate to a negligible level by tightening the $m(\Km\pip)$ range to $[1.854,1.874]\gevcc$. The \Dstarp candidates are subject to a vertex-kinematic fit which constrains the \Dstarp decay vertex to the measured position of the \epem interaction region. Only candidates having fit $\chi^2$ probabilities larger than 0.001 are retained. We retain all candidates in the less than $1\%$ of the events where multiple \Dstarp candidates are reconstructed.

\section{Kinematic weighting\label{sec:equalization}}
\begin{figure*}[ht]
\centering
%
%
%
%
%
%
%
%
%
%
%
%
%
%
%
%
%
%
\includegraphics[width=0.4\textwidth]{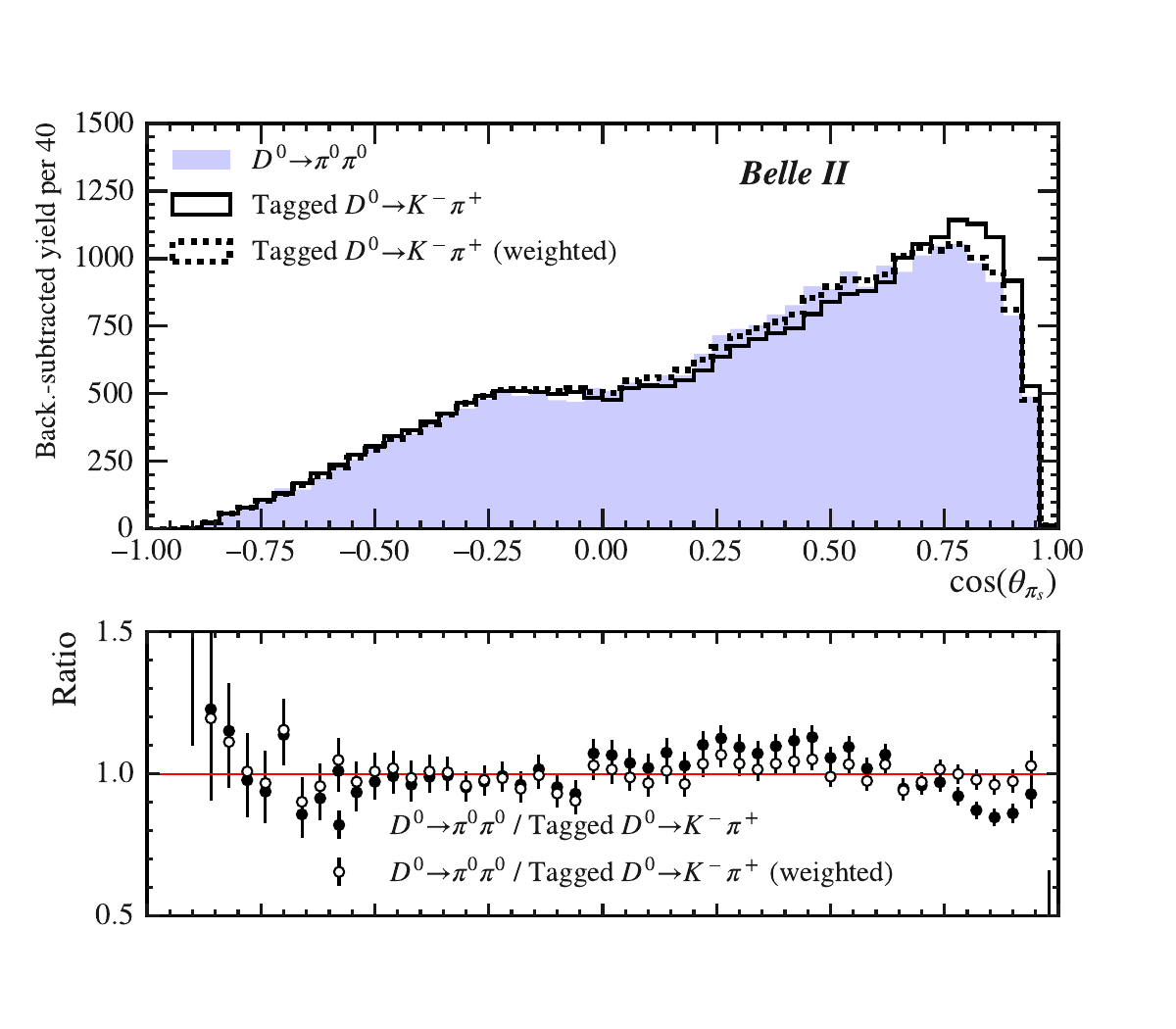}
\includegraphics[width=0.4\textwidth]{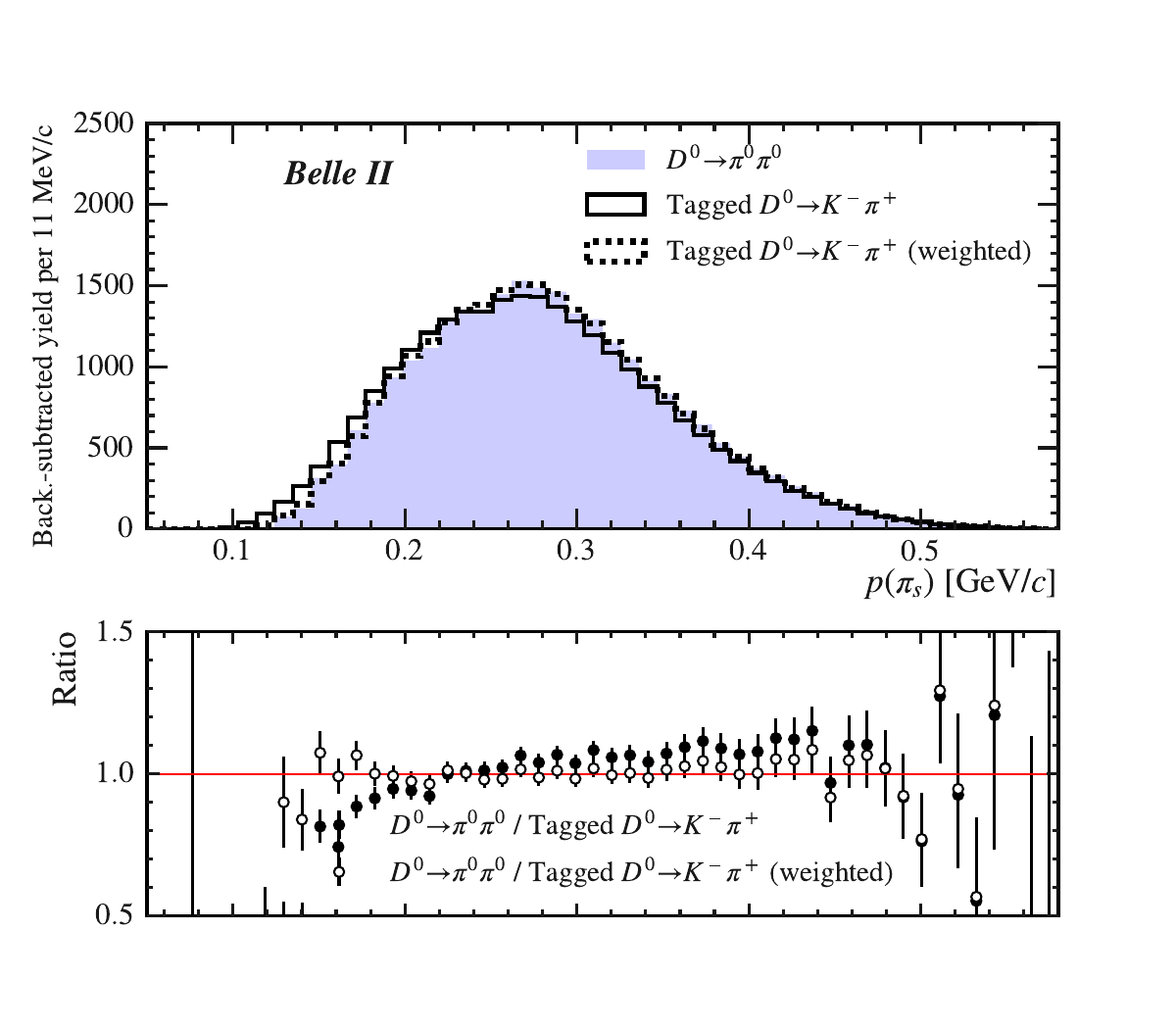}
\includegraphics[width=0.4\textwidth]{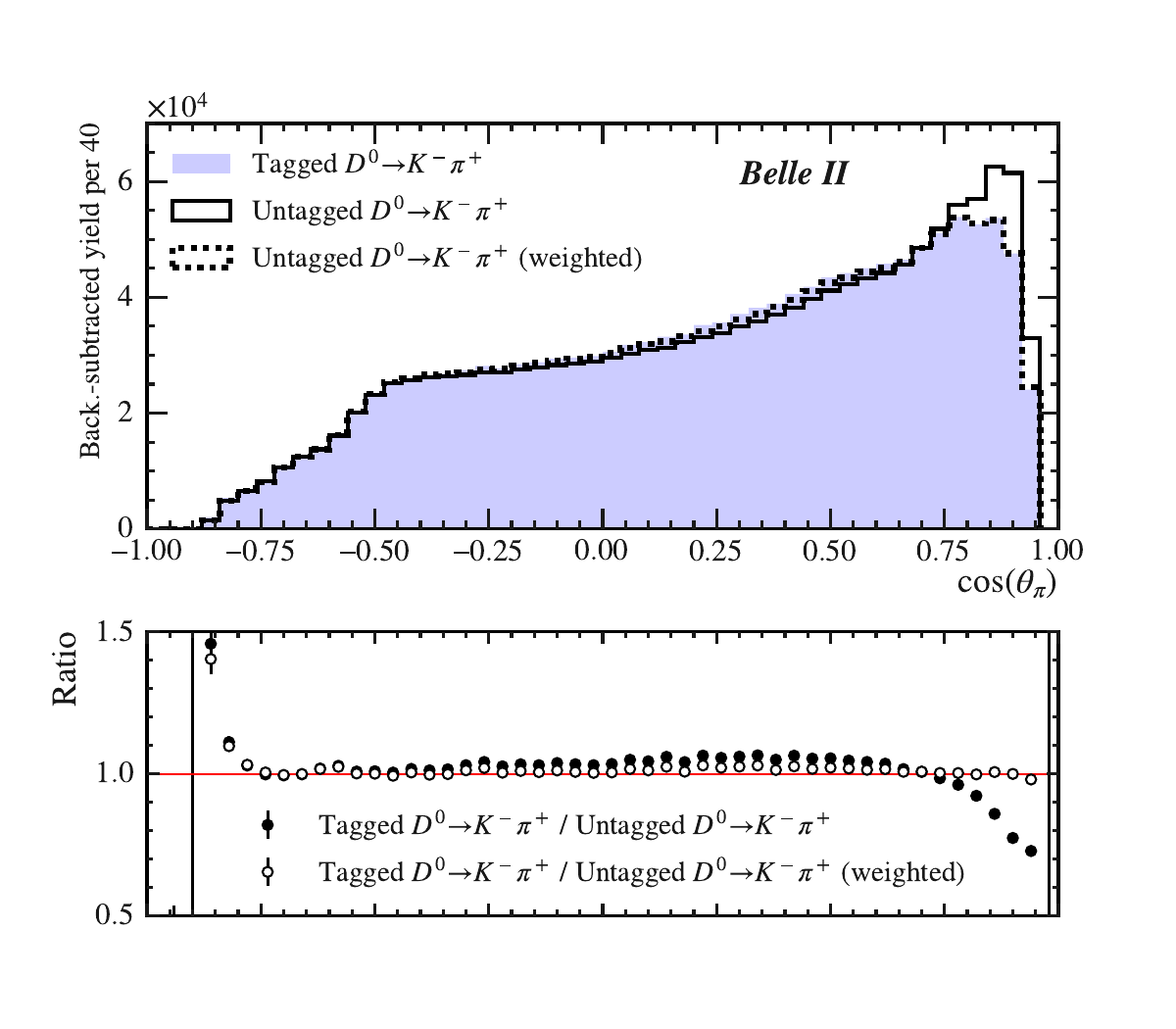}
\includegraphics[width=0.4\textwidth]{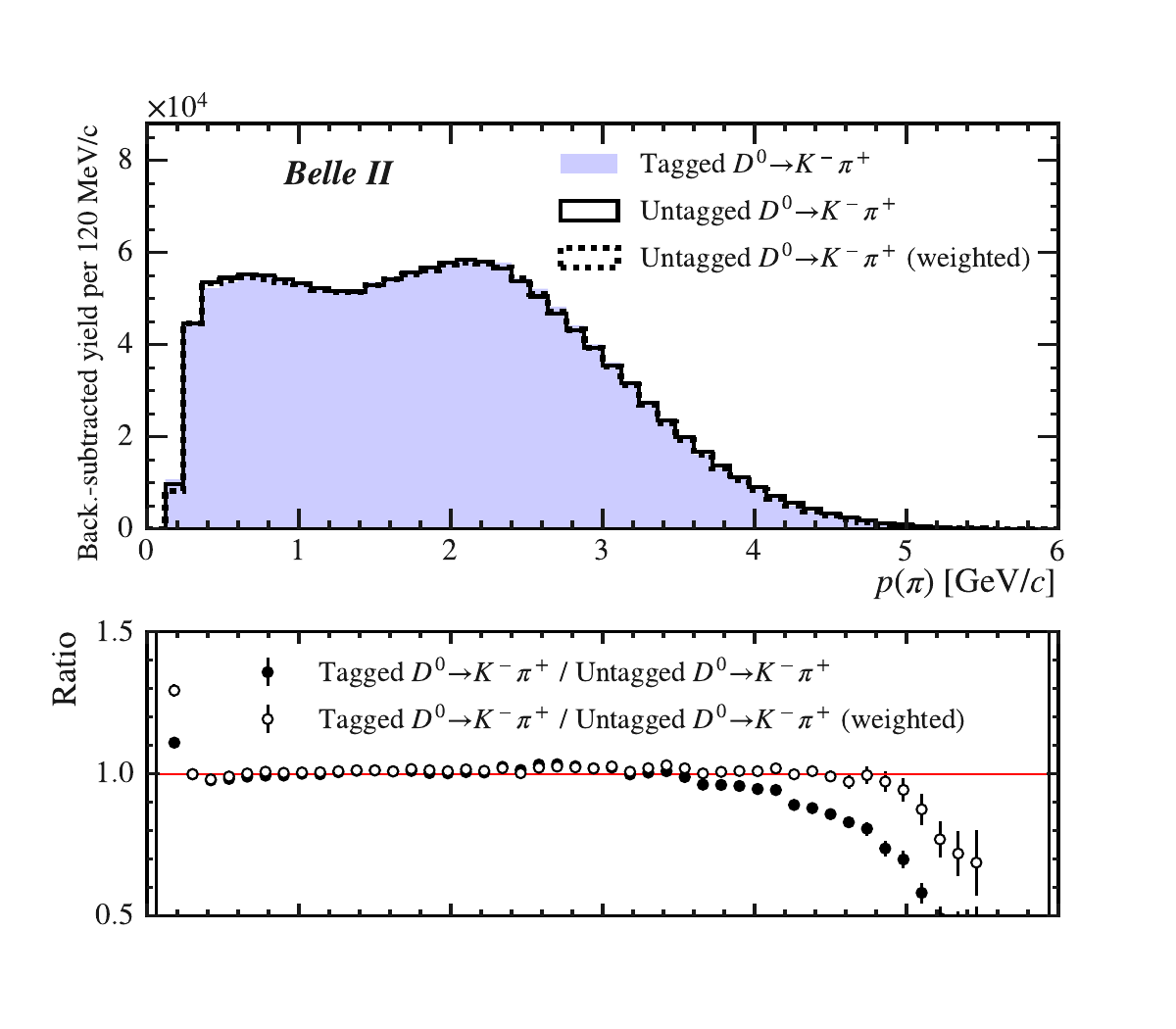}
\includegraphics[width=0.4\textwidth]{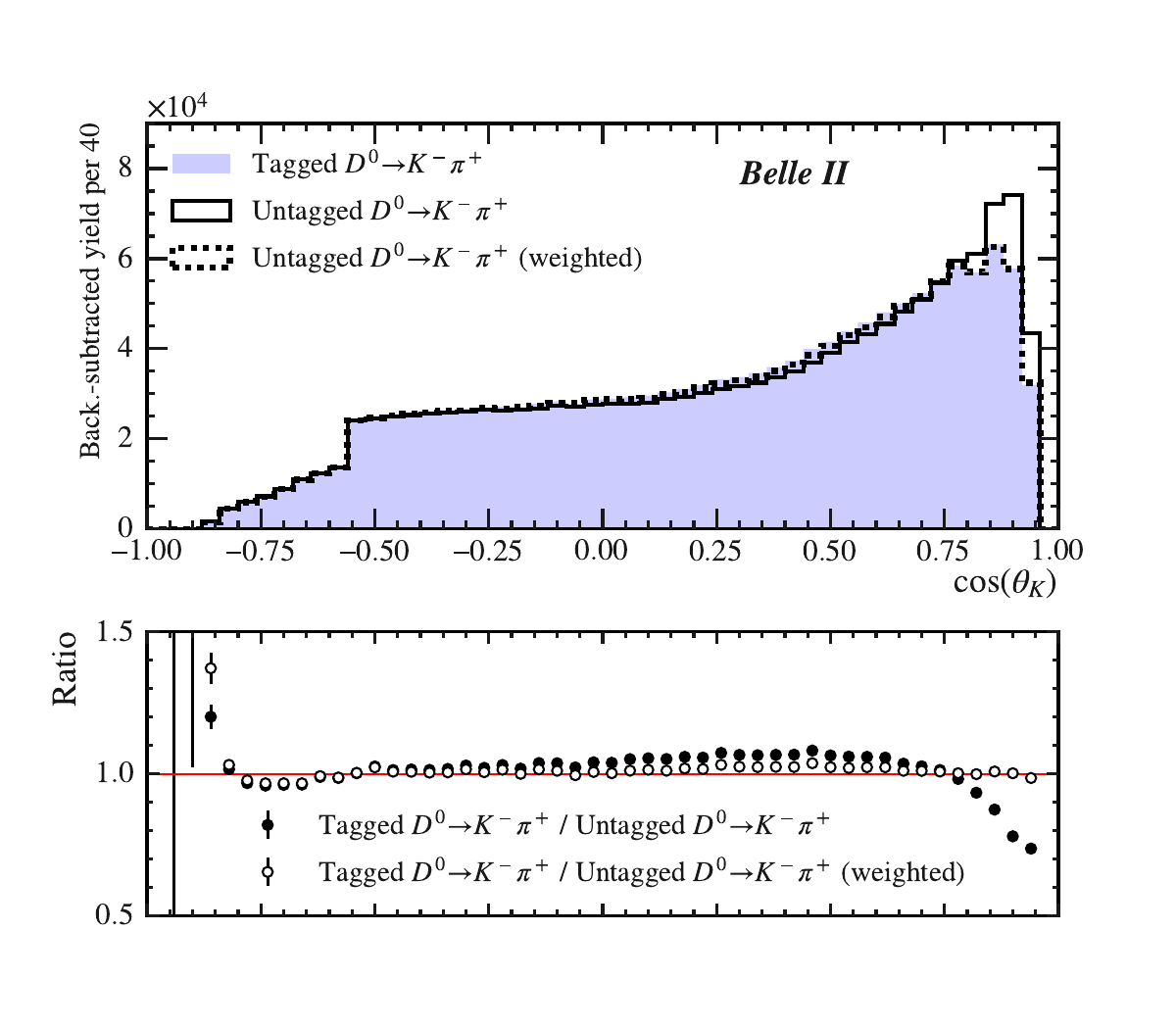}
\includegraphics[width=0.4\textwidth]{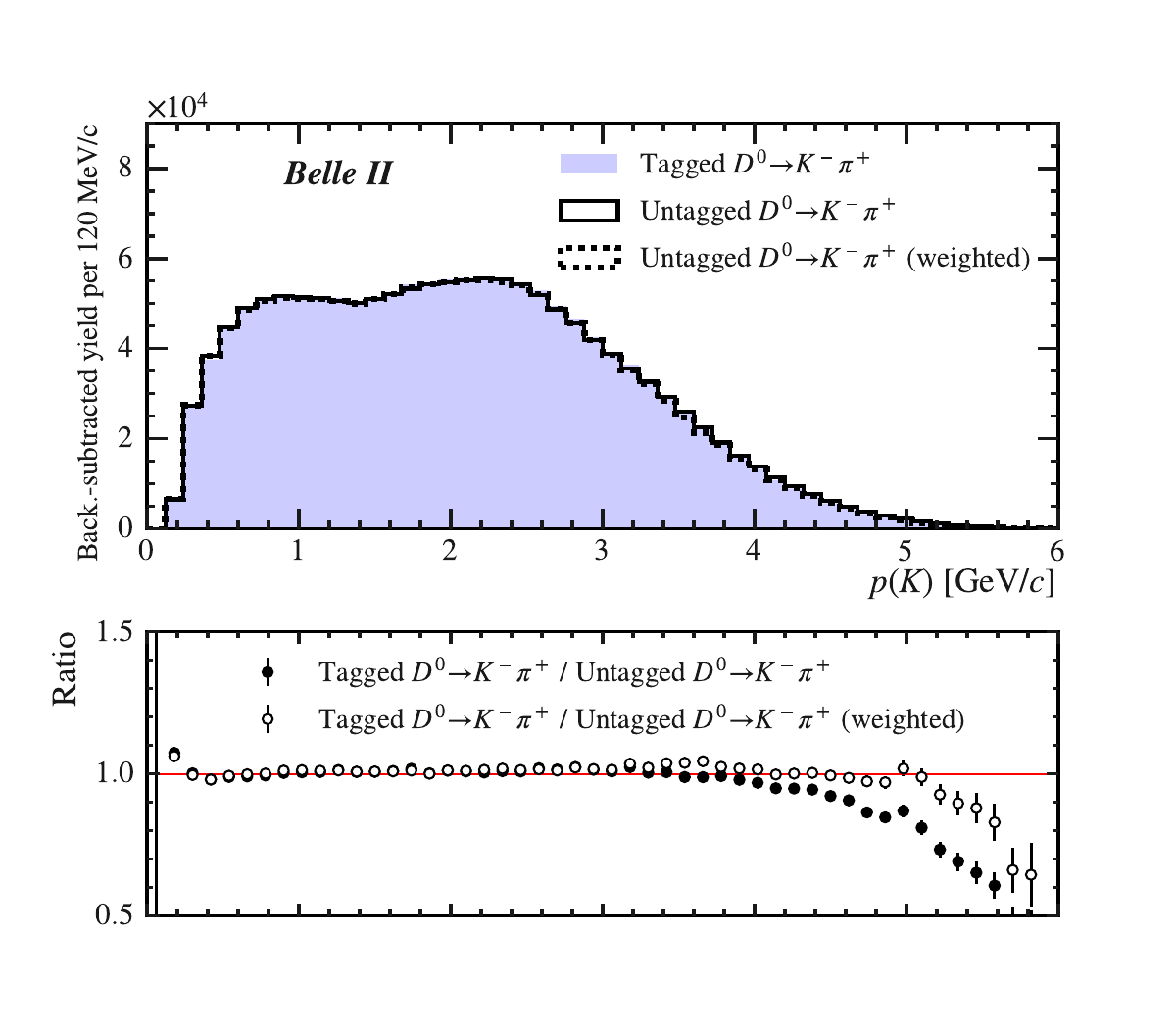}
\caption{Distributions of $\cos\theta(\pis)$ and $p(\pis)$ of background-subtracted signal and tagged control decays (first row) ; $\cos\theta(\pi)$, $p(\pi)$, $\cos\theta(K)$, $p(K)$ of background-subtracted tagged and untagged control decays (second and third row); and the ratios before and after the kinematic weighting.\label{fig:kinematic-weighting}}
\end{figure*}

Because detection asymmetries depend on kinematic properties, the cancellation provided by \cref{eq:diff_acp} is realized accurately only if the kinematic distributions across the three samples are the same. The small differences that are present are reduced using a two-step weighting procedure. In the first step, we ensure that the soft-pion detection asymmetry is accurately subtracted by matching the two-dimensional $\cos\theta(\pis)$, $p(\pis)$ distribution of the tagged control sample to that of the signal decays. In the second step, to precisely subtract the $\Km\pip$ detection asymmetry, we match the four-dimensional distribution of $\cos\theta(K)$, $p(K)$, $\cos\theta(\pi)$, $p(\pi)$ of the untagged sample to match that of the previously weighted tagged sample. The weighting steps are performed directly in two and four dimensions, so any correlations between kinematic variables are taken into account. The weights are determined on a per-candidate basis using background-subtracted signal and control decays in data with a tool that relies on boosted decision trees~\cite{Rogozhnikov:2016bdp}. The background is subtracted using the \sPlot\ method~\cite{Pivk:2004ty} and the same fit model as used for the determination of the asymmetries (\cref{sec:fit}). The effect of the weighting procedure is shown in \cref{fig:kinematic-weighting}. Small differences remain only at the borders of the distributions, and thus affect only a small subset of the three samples. Systematic uncertainties are assigned in \cref{sec:systematics} due to the imperfections in the weighting procedure.

\section{Determination of the raw asymmetries\label{sec:fit}}
The raw asymmetries are determined using unbinned maximum-likelihood fits to \Dz and \Dzb candidates in data using observables that discriminate signal and control decays from background. The fits are performed independently for positive and negative $\cos\theta_{\rm cms}$. Using ensembles of pseudoexperiments generated from the assumed probability density functions (PDFs), the fits are shown to return unbiased determinations of the asymmetries and uncertainties with proper statistical coverage.

\subsection{$\Dz\to \piz\piz$ signal sample}
In the signal sample, we use the two-dimensional $m(\piz\piz)$ \vs $\Delta m$ distribution to discriminate among four components: signal $\Dstarp\to\Dz(\to\piz\piz)\pip$ decays; background due to $\Dstarp\to\Dz(\to\KS\piz)\pip$ decays with $\KS\to\piz\piz$, in which one \piz meson from the \KS decay is not reconstructed; random-pion background from correctly reconstructed $\Dz\to\piz\piz$ decays associated with unrelated \pis candidates; and combinatorial background due to unrelated combinations of final-state particles. Signal decays peak at the expected value of the \Dz mass in $m(\piz\piz)$ and of the \Dstarp-\Dz mass difference in $\Delta m$. Candidates $\Dstarp\to\Dz(\to\KS\piz)\pip$ decays have low $m(\piz\piz)$ because of the missing \piz, but still peak in $\Delta m$ although with broader resolution than the signal. The random-pion background peaks like the signal in $m(\piz\piz)$ and does not peak in $\Delta m$. The combinatorial background does not peak in either $m(\piz\piz)$ or $\Delta m$.

The two-dimensional PDFs of signal, random-pion, and combinatorial background decays factorize into the product of two one-dimensional PDFs. The $m(\piz\piz)$ distribution of signal decays is modeled using the sum of a Johnson's $S_U$ distribution~\cite{johnson},
\begin{equation}
J(x|\mu,\sigma,\delta,\gamma) \propto \frac{e^{-\frac{1}{2}\left[ \gamma + \delta \sinh^{-1} \left( \frac{x-\mu}{\sigma} \right) \right]^2}}{\sqrt{1+\left( \frac{x-\mu}{\sigma} \right)^2}}\,,
\end{equation}
and a Gaussian function. The $\Delta m$ distribution is parameterized using a sum of a Johnson's $S_U$ and two Gaussian functions. The random-pion background shares the same $m(\piz\piz)$ PDF as the signal and has a $\Delta m$ distribution modeled by $(\Delta m - m_{\pip})^{\frac{1}{2}} + \alpha (\Delta m - m_{\pip})^{\frac{3}{2}} + \beta (\Delta m - m_{\pip})^{\frac{5}{2}}$, with $m_{\pip}$ being the known value of the charged-pion mass~\cite{pdg}. The combinatorial background shares the same $\Delta m$ distribution used for random-pion decays and has $m(\piz\piz)$ modeled by a second-order polynomial.

In the $\Dz\to\KS\piz$ component, the width of the $\Delta m$ distribution correlates with $m(\piz\piz)$, which we parametrize analytically as a second-order polynomial. The parameters of the polynomial are determined from simulation. The two-dimensional PDF is the product of the $\Delta m$ PDF, conditional on the value of $m(\piz\piz)$, and of the $m(\piz\piz)$ PDF. The first term is parametrized as the sum of a Johnson's $S_U$ function and a Gaussian function where the $\sigma$ parameter of the former accounts for the aforementioned correlation. The second term is given by the sum of a Gaussian and an exponential function.

Omitting the fit parameters from the list of arguments to simplify the notation, the total PDF is
\begin{equation}\label{eq:signal_pdf_full}
P(m,\Delta m|...) = \sum_i f^i (1+qA^{i}) P^i(m,\Delta m|...)\,,
\end{equation}
where $q=1(-1)$ for \Dz (\Dzb) candidates, and $f^i$ and $A^i$ are the fraction and raw asymmetry of the component $i$ (and one fraction is expressed in terms of the others for proper normalization of the PDF). When fitting to the data, 18 parameters are floated: the component fractions and asymmetries, the $\mu$ and $\gamma$ parameters of the Johnson's $S_U$ function of the signal $m(\piz\piz)$ and $\Delta m$ PDFs, the mean of the Gaussian function of the signal $m(\piz\piz)$ PDF, the $\mu$ and $\sigma$ of the Johnson's $S_U$ function of the $\Dz\to\KS\piz$ $\Delta m$ PDF, the shape parameters of the combinatorial-background PDFs in $m(\piz\piz)$ and $\Delta m$. The remaining 21 shape parameters are fixed to the values obtained when fitting to simulated data.

\begin{figure*}[ht]
\centering
%
%
%
%
%
%
%
%
%
%
%
%
\includegraphics[width=0.4\textwidth]{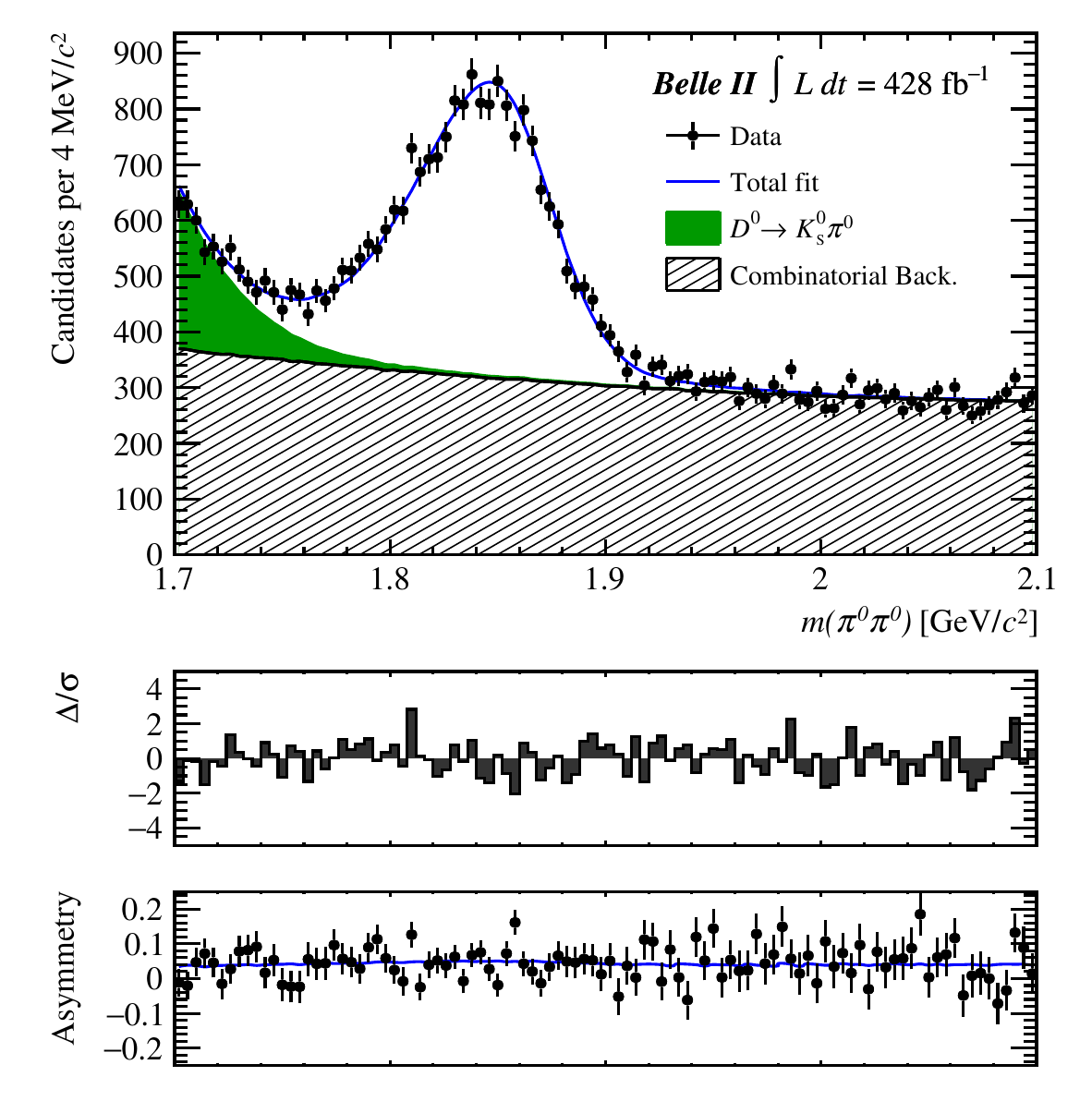}
\includegraphics[width=0.4\textwidth]{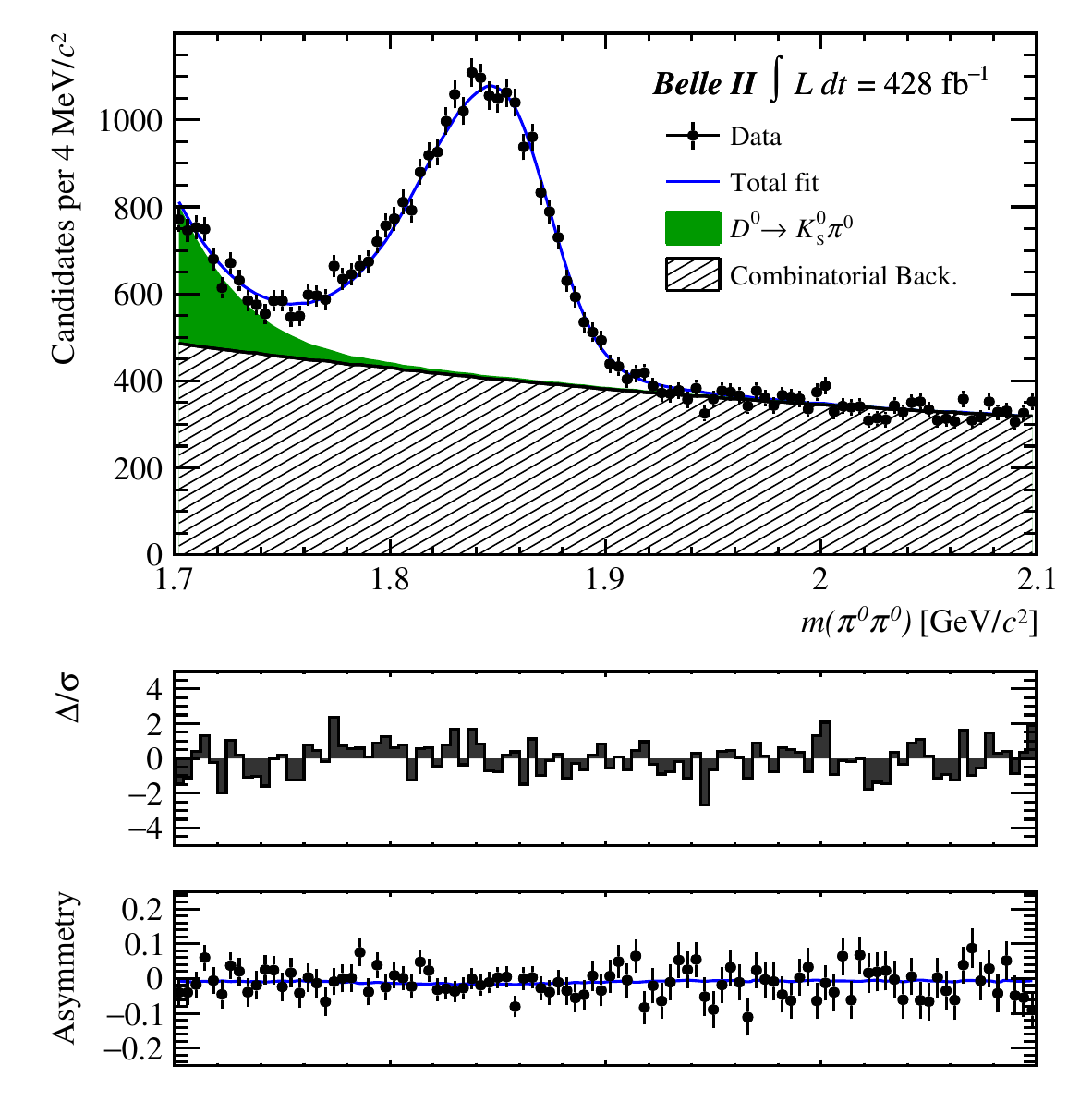}
\includegraphics[width=0.4\textwidth]{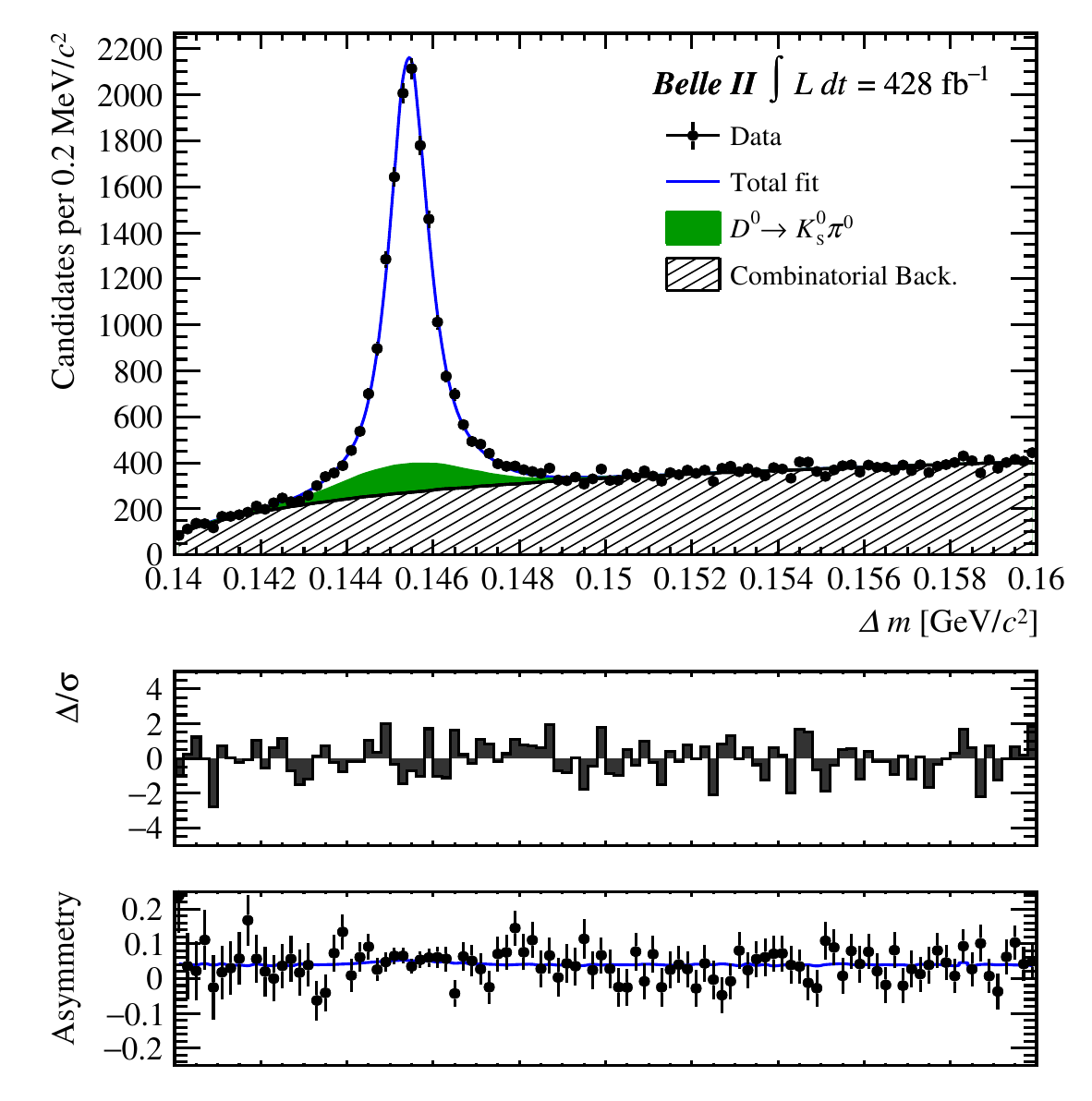}
\includegraphics[width=0.4\textwidth]{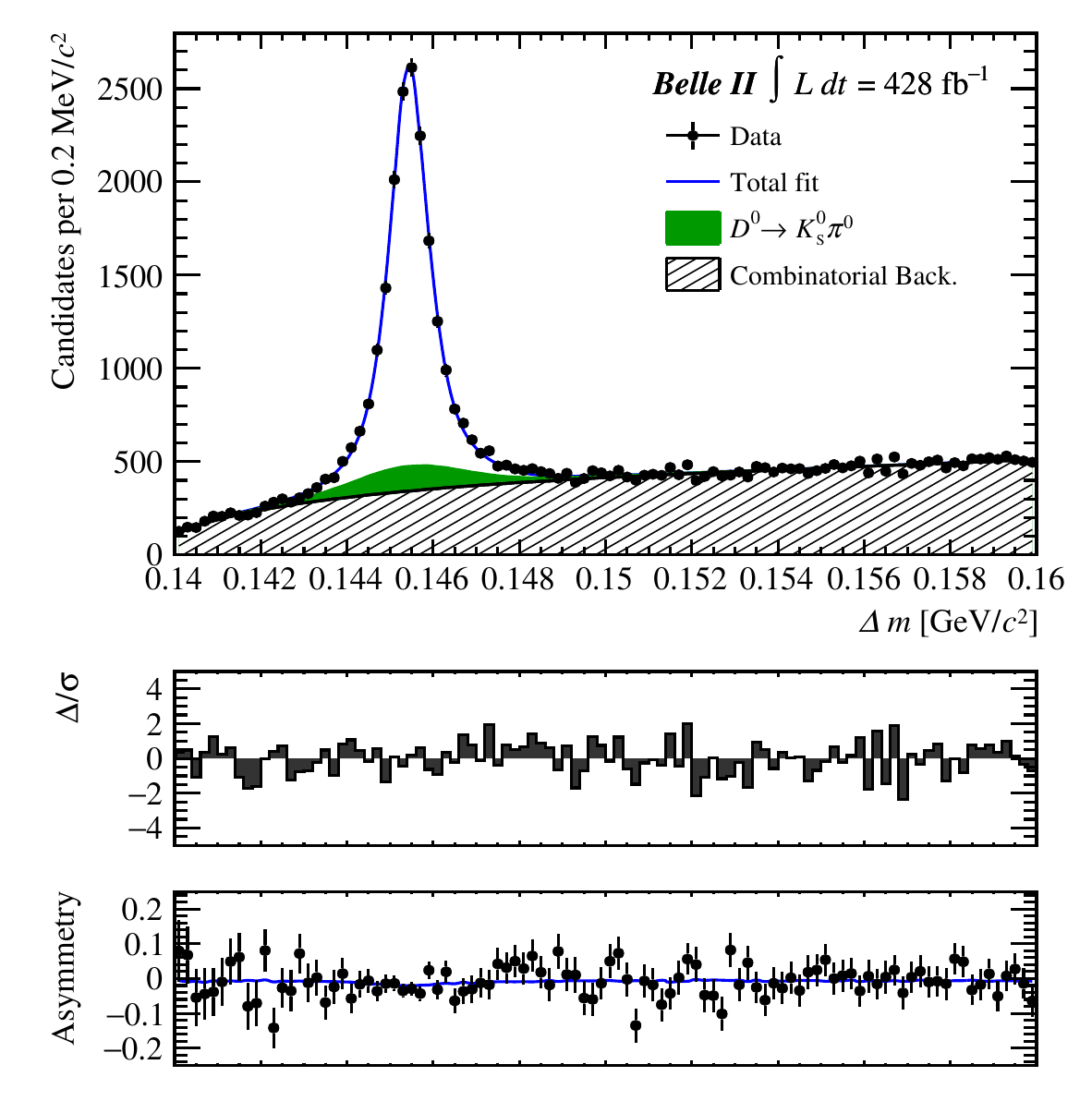}
\caption{Distributions of $m(\piz\piz)$ (top) and $\Delta m$ (bottom) for $\Dz\to\piz\piz$ candidates with $\cos\theta_{\rm cms}(\Dstarp)<0$ (left) and $\cos\theta_{\rm cms}(\Dstarp)>0$ (right), with fit projections overlaid. The middle panel of each plot shows the distribution of the difference between observed and fit values divided by the uncertainty (pull), the bottom panel shows the asymmetry between \Dz and \Dzb candidates with the fit projection overlaid.}\label{fig:signal_fit_data_b2}
\end{figure*}

\Cref{fig:signal_fit_data_b2} shows the $m(\piz\piz)$ and $\Delta m$ distributions of the data split in the positive and negative $\cos\theta_{\rm cms}$, with fit projection overlaid. In this fit, the random-pion component is neglected because its fraction is found to be consistent with zero (removing two floating parameters from the fit). The fit model describes the data well and the signal yields are measured to be $14\,100\pm130$ and $11\,550\pm110$ in the forward and backward bins, respectively. The corresponding raw asymmetries are $(-2.20\pm0.95)\%$ and $(5.66\pm1.05)\%$. When averaged we obtain
\begin{equation}
A^{\prime\,\piz\piz} = (1.73\pm0.71)\%\,.
\end{equation}
The uncertainties are statistical only.

\subsection{$\Dz\to\Km\pip$ control samples}
\begin{figure*}[ht]
%
%
%
%
%
%
%
\includegraphics[width=0.4\textwidth]{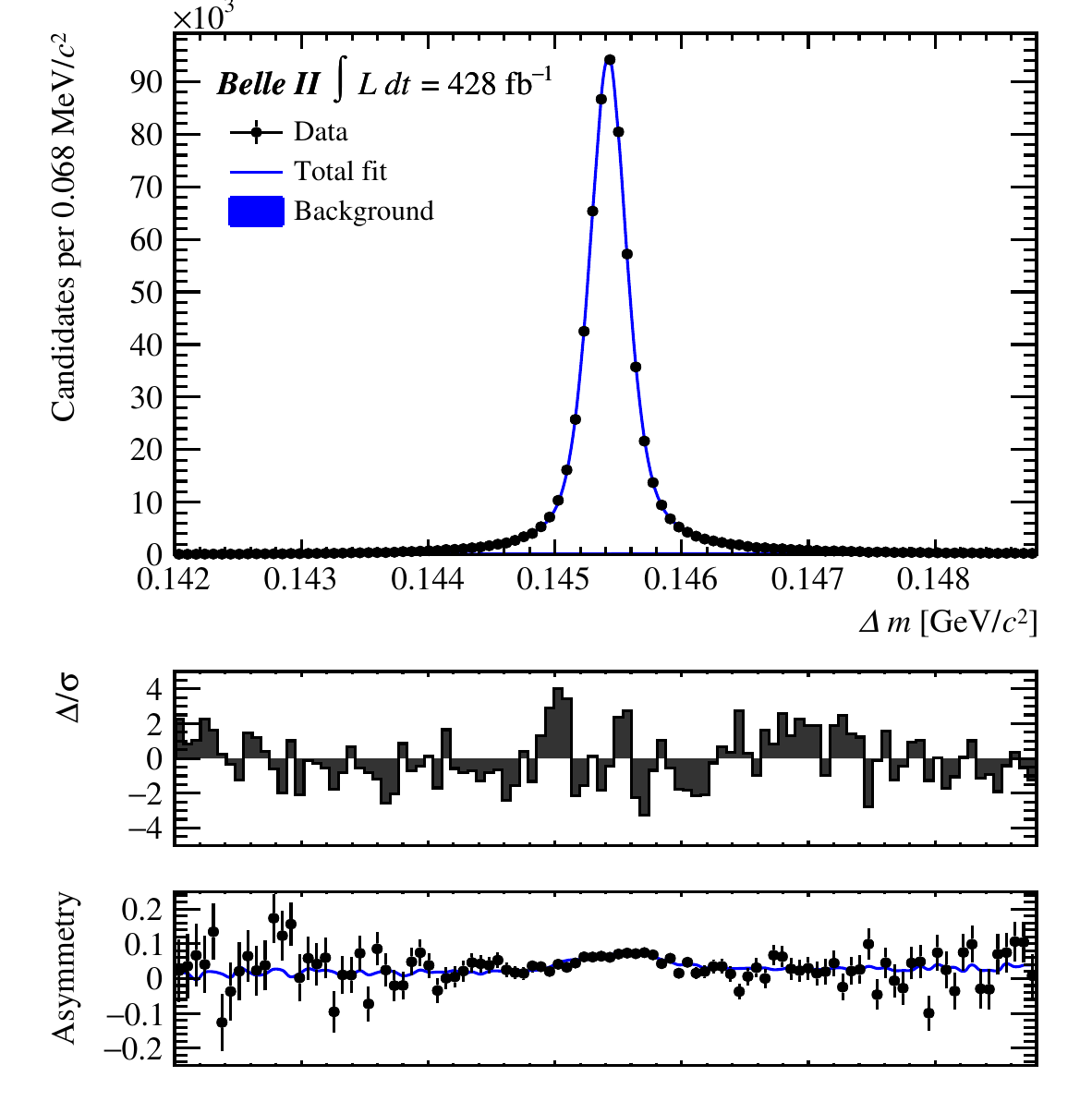}
\includegraphics[width=0.4\textwidth]{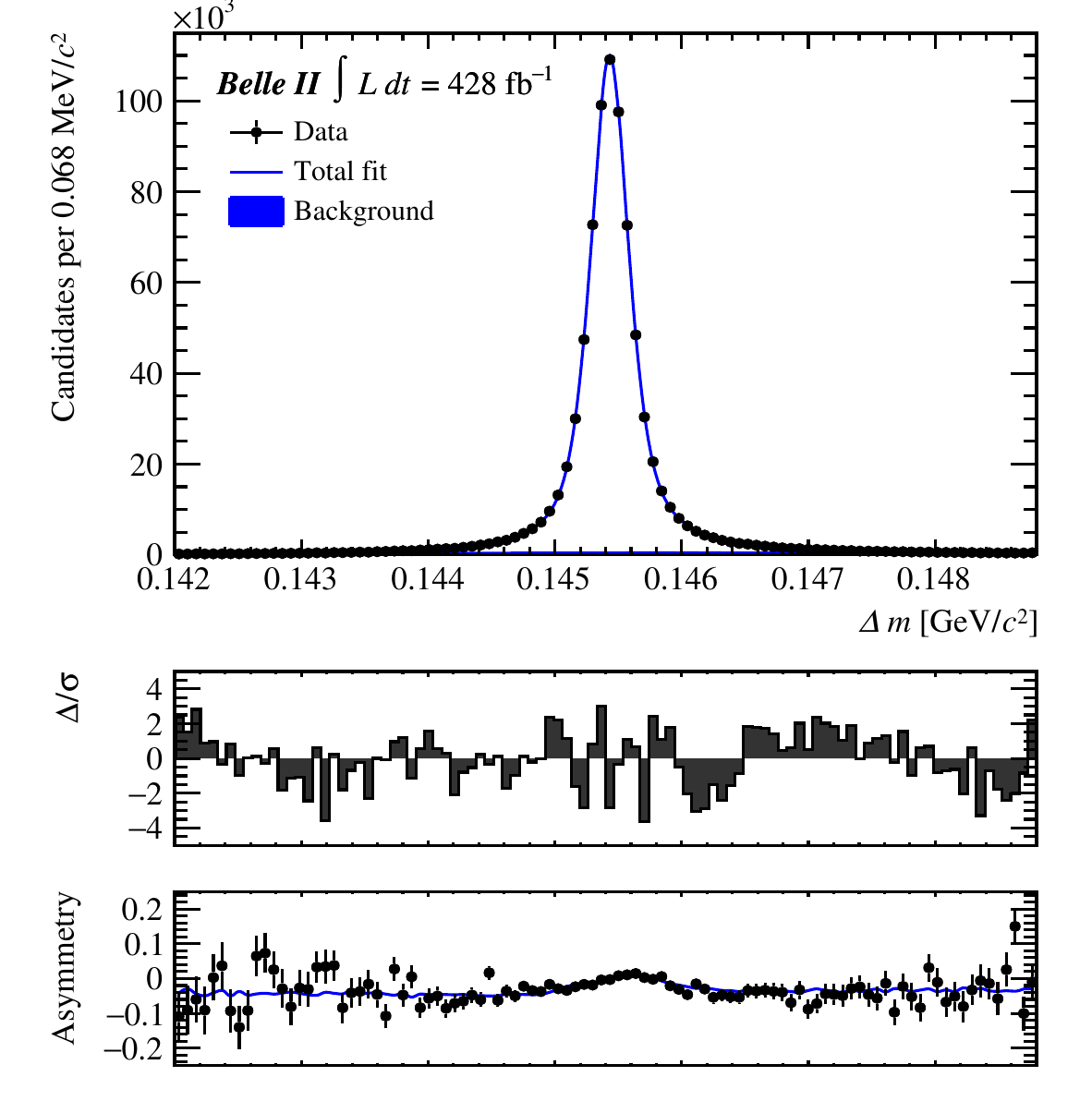}
\caption{Distributions of $\Delta m$ for tagged $\Dz\to\Km\pip$ candidates with $\cos\theta_{\rm cms}(\Dstar)<0$ (left) and $\cos\theta_{\rm cms}(\Dstar)>0$ (right), with fit projections overlaid. The middle panel of each plot shows the distribution of the difference between observed and fit values divided by the uncertainty (pull), the bottom panel shows the asymmetry between \Dz and \Dzb candidates with the fit projection overlaid.}\label{fig:control_fit_deltam_data}
\end{figure*}

The raw asymmetry of the tagged control sample is determined from a fit to the $\Delta m$ distribution, in which only two components are considered: the $\Dstarp\to\Dz(\to\Km\pip)\pip$ decays, and a background made of both random-pion and combinatorial candidates. The $\Delta m$ PDF of $\Dz\to\Km\pip$ decays is the sum of a Johnson's $S_U$ and a Gaussian function, with common location and width parameters that are determined independently for \Dz and \Dzb decays to account for flavor-dependent mass biases and resolutions. The background component is modeled as $(\Delta m - m_{\pip})^{\beta} e^{-\lambda(\Delta m- m_{\pip})}$. The relative fractions of the components, their asymmetries and all shape parameters are floated in the fit (for a total of 14 floating parameters). \Cref{fig:control_fit_deltam_data} shows the results of the fits to the data. The measured yields of $\Dstarp\to\Dz(\to\Km\pip)\pip$ decays in the forward and backward bins are $796\,000\pm1\,200$ and $633\,700\pm1\,200$, respectively. The corresponding asymmetries, $(-0.86\pm0.13)\%$ and $(5.83\pm0.13)\%$, are averaged to obtain
\begin{equation}
A^{\prime\,K\pi} = (2.49\pm0.09)\%\,,
\end{equation}
where the uncertainties are statistical only.

\begin{figure*}[ht]
\centering
%
%
%
%
%
%
\includegraphics[width=0.4\textwidth]{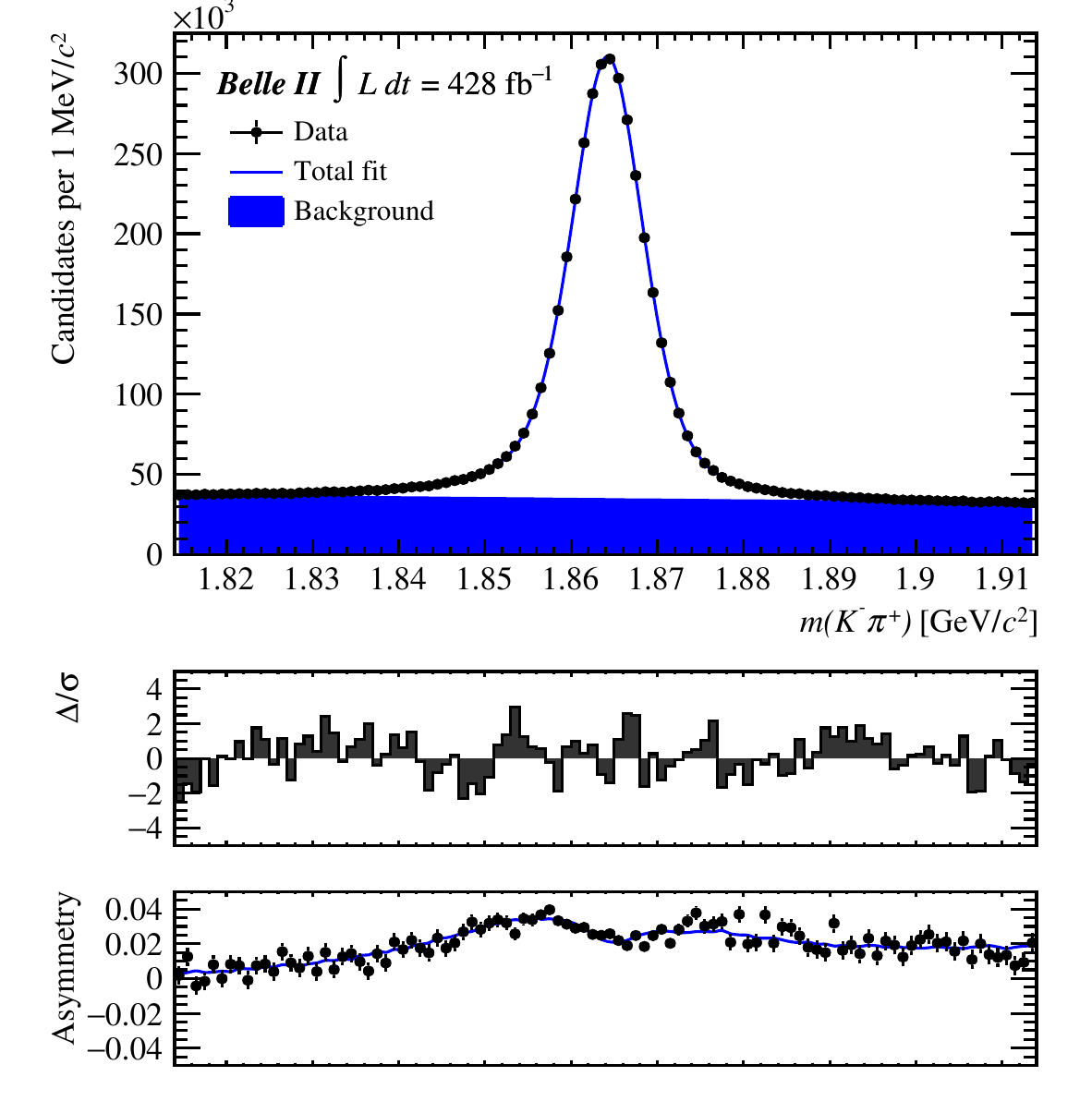}
\includegraphics[width=0.4\textwidth]{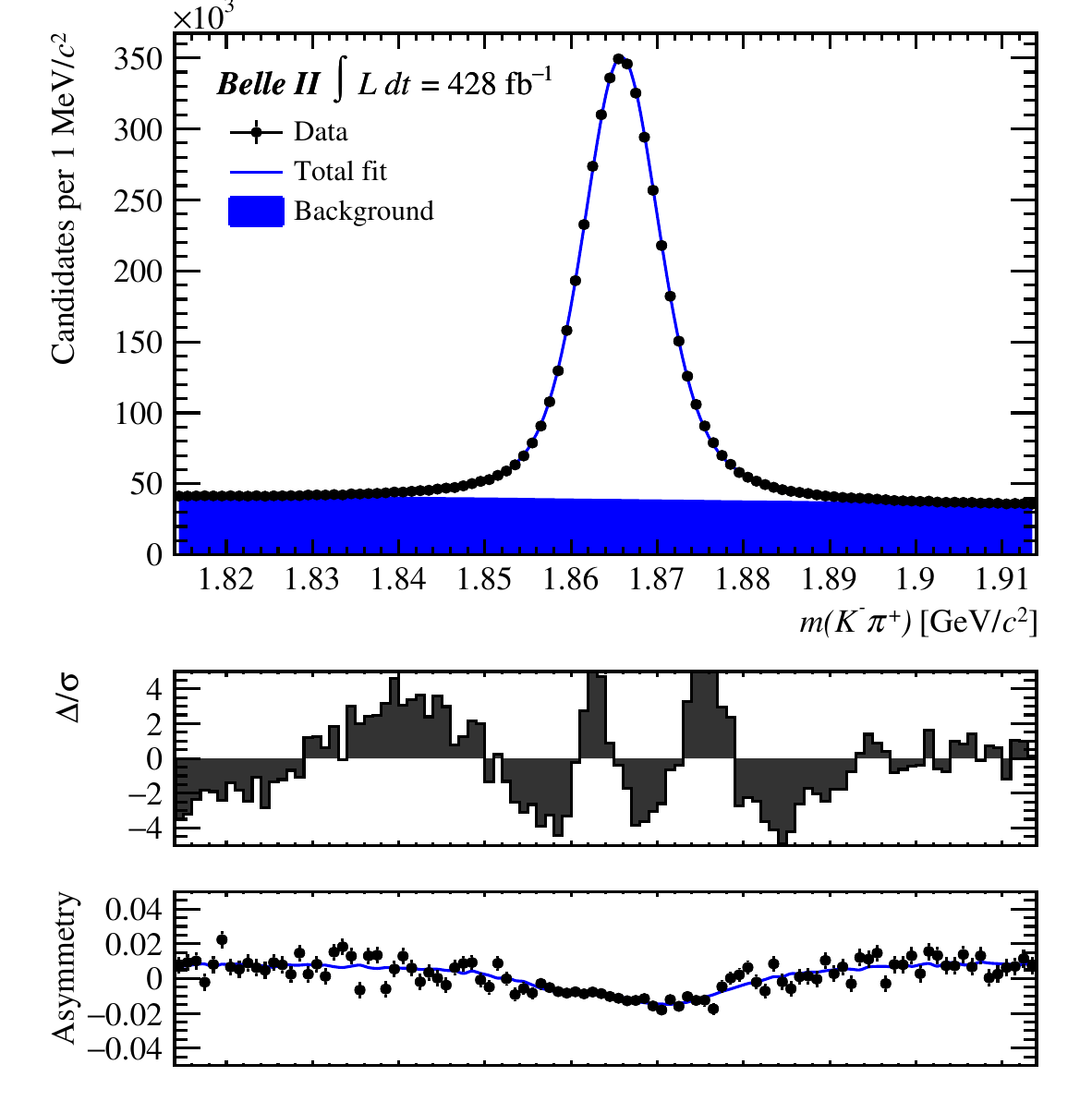}
\caption{Distributions of $m(\Km\pip)$ for untagged $\Dz\to\Km\pip$ candidates with $\cos\theta_{\rm cms}(\Dz)<0$ (left) and $\cos\theta_{\rm cms}(\Dz)>0$ (right) in data, with fit projections overlaid. The middle panel of each plot shows the distribution of the difference between observed and fit values divided by the uncertainty (pull), the bottom panel shows the asymmetry between \Dz and \Dzb candidates with the fit projection overlaid.}\label{fig:control_fit_m_data}
\end{figure*}

For untagged decays, we fit to the $m(\Km\pip)$ distribution, again considering only two components. The untagged $\Dz\to\Km\pip$ decays are modeled using the sum of a Johnson $S_U$ and a Gaussian function, with common mean and flavor-dependent width parameters. A straight line is used to model the $m(\Km\pip)$ distribution of the background with a flavor-dependent coefficient. Also in this case, all parameters are determined from the fit. The results of the fit are shown in \cref{fig:control_fit_m_data}. We measure $4\,294\,700\pm5\,600$ and $3\,374\,100\pm6\,500$ untagged $\Dz\to\Km\pip$ decays in the forward and backward bins, respectively. The corresponding asymmetries, $(-1.65\pm0.09)\%$ and $(3.75\pm0.11)\%$, are averaged to obtain
\begin{equation}
A^{\prime\,K\pi,\text{untag}} = (1.05\pm0.07)\%\,,
\end{equation}
where the uncertainties are statistical only.

The raw-asymmetry values are consistent with expected differences in reconstruction asymmetries for charged particles in forward and backward directions. While the variation of the asymmetry as a function of mass is fairly well described by the fit model for $\Dz\to\Km\pip$ control samples (\cref{fig:control_fit_m_data}, bottom panel), the \CP-averaged distributions are not (\cref{fig:control_fit_m_data}, middle panel). Systematic uncertainties are assigned in \cref{sec:systematics} to cover the impact of the mismodeling on the measured asymmetries.

Using \cref{eq:diff_acp}, we determine the time-integrated \CP asymmetry in $\Dz\to\piz\piz$ decays to be $(0.30\pm 0.72)\%$, where the uncertainty is statistical only.

\section{Systematic uncertainties\label{sec:systematics}}
The measurement is affected by the following main sources of systematic uncertainties: PDF modeling in the signal and control-sample fits, and kinematic weighting of control and signal samples. Other effects, such as those due to finite $\cos\theta_{\rm cms}$ resolution and nonuniform efficiency variation as a function of $|\cos\theta_{\rm cms}|$, are also investigated and found to be negligible. A summary of the estimated uncertainties is reported in \cref{tab:syst_sum} together with the sum in quadrature of all contributions.

\begin{table}[h]
    \caption{Summary of uncertainties affecting the measurement of $\Acp(\Dz\to\piz\piz)$. The statistical uncertainty includes contributions from the signal and the control modes.}
    \label{tab:syst_sum}
    \centering
    \begin{tabular}{lc}
    \hline\hline
    Source & Uncertainty (\%) \\
    \hline
    Modeling in the $\Dz\to\piz\piz$ fit & 0.15 \\
    Modeling in the tagged $\Dz\to\Km\pip$ fit & 0.05 \\ 
    Modeling in the untagged $\Dz\to\Km\pip$ fit & 0.09 \\ 
    Kinematic weighting & 0.09 \\
    \hline
    Total systematic & 0.20 \\
    Statistical & 0.72 \\
    \hline\hline
    \end{tabular}
\end{table}

We estimate the systematic uncertainties due to PDF modeling in the $\Dz\to\piz\piz$, tagged $\Dz\to\Km\pi$, and untagged $\Dz\to\Km\pip$ fits using simulation. From simulation we bootstrap (i.e., sampling with replacement) subsamples of the same size as the data, perform the fit, and compute the bias on the raw asymmetry as the average deviation from the true asymmetry of the full simulation sample. Fits to the simulated decays have similar qualities as observed in data. To cover possible data-simulation differences in detection and/or production asymmetries, the study is performed with input raw asymmetries for signal and control decays that span the range $[-2,7]\%$. This range includes asymmetries that are either half or double the nominal asymmetry generated by the simulation. The root-mean-squared values of the biases from the ensemble of bootstrap samples are assigned as systematic uncertainties due to the imperfect modeling. The default fit models assume that most shape parameters are flavor-independent. To verify this assumption, we refit the data replacing individual shape parameters with flavor-dependent ones. In all cases, the parameter asymmetries are consistent with zero and statistically insignificant shifts are observed in the measured \Acp values. The default $\Dz\to\piz\piz$ model includes 37 parameters, where 16 are left floating and the remaining 21 are fixed to the values obtained from simulation. The robustness of this setup is checked by varying the fixed parameters within their uncertainties. The impact on $A^{\prime\,\piz\piz}$ is found to be negligible compared to other uncertainties.

We estimate the systematic uncertainty due to the kinematic weighting by considering two sources of uncertainty: the accuracy of the background-subtraction procedure to obtain the kinematic distributions of the signal and control decays, and the choice of the variables used in the weighting procedure. For background subtraction, we rely on the \sPlot\ method. Simulation shows that this approach could introduce an absolute shift in the estimated \CP asymmetry of $0.05\%$. To test how the choice of kinematic variables impacts the result, we develop an alternative weighting procedure by replacing the momenta of the soft pion, kaon, and pion with their transverse momenta. The shift in \Acp observed when using the alternative weights is $0.07\%$. The sum in quadrature of the two shifts is assigned as systematic uncertainty due to the weighting procedure.

\section{Final results and conclusions\label{sec:results}}
Using \Dstarp-tagged $\Dz\to\piz\piz$ decays reconstructed in the data sample collected by Belle~II between 2019 and 2022, which corresponds to 428\invfb of integrated luminosity, we measure the time-integrated \CP asymmetry in $\Dz\to\piz\piz$ decays to be
\begin{equation}
\Acp(\Dz\to\piz\piz) = (0.30\pm0.72 \pm 0.20)\%\,,
\end{equation}
where the first uncertainty is statistical and the second systematic. The result is consistent with \CP symmetry and with the best existing measurement, from Belle~\cite{Nisar:2014}. It is 15\% less precise than the Belle measurement, but it is based on a data sample less than one-half the size. The improved precision per luminosity is achieved through an improved event selection, which exploits Belle~II's superior capabilities in the reconstruction of neutral pions.

We compute the isospin sum-rule of \cref{eq:sum-rule} using our result, LHCb's measurement of $A^\text{dir}_{\CP}(\Dz\to\pip\pim)$~\cite{LHCb:2022lry}, the world-average value of $A^\text{dir}_{\CP}(\Dp\to\pip\piz)$~\cite{HFLAV}, Belle's measurement of $\Acp(\Dz\to\piz\piz)$~\cite{Nisar:2014}, LHCb's measurement of the indirect \CP-violation parameter $\Delta Y$~\cite{LHCb:2021vmn}, the world-average values of the $D\to\pi\pi$ branching fractions~\cite{pdg}, and the world-average values of the \Dz and \Dp lifetimes~\cite{pdg}. The direct \CP asymmetry in $\Dz\to\piz\piz$ is obtained by subtracting $\Delta Y$ from the time-integrated asymmetry assuming universality~\cite{Grossman:2006jg}, and that the average decay time of the selected $\Dz\to\piz\piz$ decays is equal to the \Dz lifetime~\cite{CDF:2011ejf}. All inputs are assumed to be uncorrelated. We obtain $R = (1.5\pm 2.5)\times10^{-3}$, which shows that our measurement of $\Acp(\Dz\to\piz\piz)$ improves the precision of the sum rule by approximately 20\% compared to the current determination~\cite{HFLAV}. The sum rule is still limited by the precision on $\Acp(\Dz\to\piz\piz)$, so future measurements based on the larger samples being collected at Belle II will be critical to further constrain its value. %
 
%
%
This work, based on data collected using the Belle II detector, which was built and commissioned prior to March 2019,
was supported by
Higher Education and Science Committee of the Republic of Armenia Grant No.~23LCG-1C011;
Australian Research Council and Research Grants
No.~DP200101792, %
No.~DP210101900, %
No.~DP210102831, %
No.~DE220100462, %
No.~LE210100098, %
and
No.~LE230100085; %
Austrian Federal Ministry of Education, Science and Research,
Austrian Science Fund (FWF) Grants
DOI:~10.55776/P34529,
DOI:~10.55776/J4731,
DOI:~10.55776/J4625,
DOI:~10.55776/M3153,
and
DOI:~10.55776/PAT1836324,
and
Horizon 2020 ERC Starting Grant No.~947006 ``InterLeptons'';
Natural Sciences and Engineering Research Council of Canada, Compute Canada and CANARIE;
National Key R\&D Program of China under Contract No.~2024YFA1610503,
and
No.~2024YFA1610504
National Natural Science Foundation of China and Research Grants
No.~11575017,
No.~11761141009,
No.~11705209,
No.~11975076,
No.~12135005,
No.~12150004,
No.~12161141008,
No.~12475093,
and
No.~12175041,
and Shandong Provincial Natural Science Foundation Project~ZR2022JQ02;
the Czech Science Foundation Grant No.~22-18469S 
and
Charles University Grant Agency project No.~246122;
European Research Council, Seventh Framework PIEF-GA-2013-622527,
Horizon 2020 ERC-Advanced Grants No.~267104 and No.~884719,
Horizon 2020 ERC-Consolidator Grant No.~819127,
Horizon 2020 Marie Sklodowska-Curie Grant Agreement No.~700525 ``NIOBE''
and
No.~101026516,
and
Horizon 2020 Marie Sklodowska-Curie RISE project JENNIFER2 Grant Agreement No.~822070 (European grants);
L'Institut National de Physique Nucl\'{e}aire et de Physique des Particules (IN2P3) du CNRS
and
L'Agence Nationale de la Recherche (ANR) under Grant No.~ANR-21-CE31-0009 (France);
BMBF, DFG, HGF, MPG, and AvH Foundation (Germany);
Department of Atomic Energy under Project Identification No.~RTI 4002,
Department of Science and Technology,
and
UPES SEED funding programs
No.~UPES/R\&D-SEED-INFRA/17052023/01 and
No.~UPES/R\&D-SOE/20062022/06 (India);
Israel Science Foundation Grant No.~2476/17,
U.S.-Israel Binational Science Foundation Grant No.~2016113, and
Israel Ministry of Science Grant No.~3-16543;
Istituto Nazionale di Fisica Nucleare and the Research Grants BELLE2,
and
the ICSC – Centro Nazionale di Ricerca in High Performance Computing, Big Data and Quantum Computing, funded by European Union – NextGenerationEU;
Japan Society for the Promotion of Science, Grant-in-Aid for Scientific Research Grants
No.~16H03968,
No.~16H03993,
No.~16H06492,
No.~16K05323,
No.~17H01133,
No.~17H05405,
No.~18K03621,
No.~18H03710,
No.~18H05226,
No.~19H00682, %
No.~20H05850,
No.~20H05858,
No.~22H00144,
No.~22K14056,
No.~22K21347,
No.~23H05433,
No.~26220706,
and
No.~26400255,
and
the Ministry of Education, Culture, Sports, Science, and Technology (MEXT) of Japan;  
National Research Foundation (NRF) of Korea Grants
No.~2016R1-D1A1B-02012900,
No.~2018R1-A6A1A-06024970,
No.~2021R1-A6A1A-03043957,
No.~2021R1-F1A-1060423,
No.~2021R1-F1A-1064008,
No.~2022R1-A2C-1003993,
No.~2022R1-A2C-1092335,
No.~RS-2023-00208693,
No.~RS-2024-00354342
and
No.~RS-2022-00197659,
Radiation Science Research Institute,
Foreign Large-Size Research Facility Application Supporting project,
the Global Science Experimental Data Hub Center, the Korea Institute of
Science and Technology Information (K24L2M1C4)
and
KREONET/GLORIAD;
Universiti Malaya RU grant, Akademi Sains Malaysia, and Ministry of Education Malaysia;
Frontiers of Science Program Contracts
No.~FOINS-296,
No.~CB-221329,
No.~CB-236394,
No.~CB-254409,
and
No.~CB-180023, and SEP-CINVESTAV Research Grant No.~237 (Mexico);
the Polish Ministry of Science and Higher Education and the National Science Center;
the Ministry of Science and Higher Education of the Russian Federation
and
the HSE University Basic Research Program, Moscow;
University of Tabuk Research Grants
No.~S-0256-1438 and No.~S-0280-1439 (Saudi Arabia), and
Researchers Supporting Project number (RSPD2025R873), King Saud University, Riyadh,
Saudi Arabia;
Slovenian Research Agency and Research Grants
No.~J1-50010
and
No.~P1-0135;
Ikerbasque, Basque Foundation for Science,
State Agency for Research of the Spanish Ministry of Science and Innovation through Grant No. PID2022-136510NB-C33, Spain,
Agencia Estatal de Investigacion, Spain
Grant No.~RYC2020-029875-I
and
Generalitat Valenciana, Spain
Grant No.~CIDEGENT/2018/020;
The Knut and Alice Wallenberg Foundation (Sweden), Contracts No.~2021.0174 and No.~2021.0299;
National Science and Technology Council,
and
Ministry of Education (Taiwan);
Thailand Center of Excellence in Physics;
TUBITAK ULAKBIM (Turkey);
National Research Foundation of Ukraine, Project No.~2020.02/0257,
and
Ministry of Education and Science of Ukraine;
the U.S. National Science Foundation and Research Grants
No.~PHY-1913789 %
and
No.~PHY-2111604, %
and the U.S. Department of Energy and Research Awards
No.~DE-AC06-76RLO1830, %
No.~DE-SC0007983, %
No.~DE-SC0009824, %
No.~DE-SC0009973, %
No.~DE-SC0010007, %
No.~DE-SC0010073, %
No.~DE-SC0010118, %
No.~DE-SC0010504, %
No.~DE-SC0011784, %
No.~DE-SC0012704, %
No.~DE-SC0019230, %
No.~DE-SC0021274, %
No.~DE-SC0021616, %
No.~DE-SC0022350, %
No.~DE-SC0023470; %
and
the Vietnam Academy of Science and Technology (VAST) under Grants
No.~NVCC.05.12/22-23
and
No.~DL0000.02/24-25.

These acknowledgements are not to be interpreted as an endorsement of any statement made
by any of our institutes, funding agencies, governments, or their representatives.

We thank the SuperKEKB team for delivering high-luminosity collisions;
the KEK cryogenics group for the efficient operation of the detector solenoid magnet and IBBelle on site;
the KEK Computer Research Center for on-site computing support; the NII for SINET6 network support;
and the raw-data centers hosted by BNL, DESY, GridKa, IN2P3, INFN, 
and the University of Victoria.
 
\bibliographystyle{belle2}

\begin{thebibliography}{10}

\bibitem{Golden:1989qx}
M.~Golden and B.~Grinstein,
  \ifthenelse{\boolean{articletitles}}{\emph{{Enhanced \CP violations in
  hadronic charm decays}},
  }{}\href{https://doi.org/10.1016/0370-2693(89)90353-5}{Phys.\ Lett.\ B
  \textbf{222} (1989) 501}.

\bibitem{Buccella:1994nf}
F.~Buccella {\em et~al.},
  \ifthenelse{\boolean{articletitles}}{\emph{{Nonleptonic weak decays of
  charmed mesons}}, }{}\href{https://doi.org/10.1103/PhysRevD.51.3478}{Phys.\
  Rev.\ D \textbf{51} (1995) 3478},
  \href{http://arxiv.org/abs/hep-ph/9411286}{{\normalfont\ttfamily
  arXiv:hep-ph/9411286}}.

\bibitem{Bianco:2003vb}
S.~Bianco, F.~L. Fabbri, D.~Benson, and I.~Bigi,
  \ifthenelse{\boolean{articletitles}}{\emph{{A Cicerone for the physics of
  charm}}, }{}\href{https://doi.org/10.1393/ncr/i2003-10003-1}{Riv.\ Nuovo
  Cim.\  \textbf{26N7} (2003) 1},
  \href{http://arxiv.org/abs/hep-ex/0309021}{{\normalfont\ttfamily
  arXiv:hep-ex/0309021}}.

\bibitem{Grossman:2006jg}
Y.~Grossman, A.~L. Kagan, and Y.~Nir,
  \ifthenelse{\boolean{articletitles}}{\emph{{New physics and \CP violation in
  singly Cabibbo suppressed $D$ decays}},
  }{}\href{https://doi.org/10.1103/PhysRevD.75.036008}{Phys.\ Rev.\ D
  \textbf{75} (2007) 036008},
  \href{http://arxiv.org/abs/hep-ph/0609178}{{\normalfont\ttfamily
  arXiv:hep-ph/0609178}}.

\bibitem{Artuso:2008vf}
M.~Artuso, B.~Meadows, and A.~A. Petrov,
  \ifthenelse{\boolean{articletitles}}{\emph{{Charm meson decays}},
  }{}\href{https://doi.org/10.1146/annurev.nucl.58.110707.171131}{Ann.\ Rev.\
  Nucl.\ Part.\ Sci.\  \textbf{58} (2008) 249},
  \href{http://arxiv.org/abs/0802.2934}{{\normalfont\ttfamily
  arXiv:0802.2934}}.

\bibitem{Aaij:2019kcg}
LHCb collaboration, R.~Aaij {\em et~al.},
  \ifthenelse{\boolean{articletitles}}{\emph{{Observation of \CP violation in
  charm decays}},
  }{}\href{https://doi.org/10.1103/PhysRevLett.122.211803}{Phys.\ Rev.\ Lett.\
  \textbf{122} (2019) 211803},
  \href{http://arxiv.org/abs/1903.08726}{{\normalfont\ttfamily
  arXiv:1903.08726}}.

\bibitem{LHCb:2022lry}
LHCb collaboration, R.~Aaij {\em et~al.},
  \ifthenelse{\boolean{articletitles}}{\emph{{Measurement of the
  time-integrated \CP asymmetry in $\Dz\to\Kp\Km$ decays}},
  }{}\href{https://doi.org/10.1103/PhysRevLett.131.091802}{Phys.\ Rev.\ Lett.\
  \textbf{131} (2023) 091802},
  \href{http://arxiv.org/abs/2209.03179}{{\normalfont\ttfamily
  arXiv:2209.03179}}.

\bibitem{Chala:2019}
M.~Chala, A.~Lenz, A.~V. Rusov, and J.~Scholtz,
  \ifthenelse{\boolean{articletitles}}{\emph{{$\Delta A_{CP}$ within the
  Standard Model and beyond}},
  }{}\href{https://doi.org/10.1007/JHEP07(2019)161}{JHEP \textbf{07} (2019)
  161}, \href{http://arxiv.org/abs/1903.10490}{{\normalfont\ttfamily
  arXiv:1903.10490}}.

\bibitem{Dery:2019ysp}
A.~Dery and Y.~Nir, \ifthenelse{\boolean{articletitles}}{\emph{{Implications of
  the LHCb discovery of \CP violation in charm decays}},
  }{}\href{https://doi.org/10.1007/JHEP12(2019)104}{JHEP \textbf{12} (2019)
  104}, \href{http://arxiv.org/abs/1909.11242}{{\normalfont\ttfamily
  arXiv:1909.11242}}.

\bibitem{Calibbi:2019bay}
L.~Calibbi, T.~Li, Y.~Li, and B.~Zhu,
  \ifthenelse{\boolean{articletitles}}{\emph{{Simple model for large \CP
  violation in charm decays, $B$-physics anomalies, muon $g-2$ and dark
  matter}}, }{}\href{https://doi.org/10.1007/JHEP10(2020)070}{JHEP \textbf{10}
  (2020) 070}, \href{http://arxiv.org/abs/1912.02676}{{\normalfont\ttfamily
  arXiv:1912.02676}}.

\bibitem{Grossman:2019}
Y.~Grossman and S.~Schacht, \ifthenelse{\boolean{articletitles}}{\emph{{The
  emergence of the $\Delta U=0$ rule in charm physics}},
  }{}\href{https://doi.org/10.1007/JHEP07(2019)020}{JHEP \textbf{07} (2019)
  020}, \href{http://arxiv.org/abs/1903.10952}{{\normalfont\ttfamily
  arXiv:1903.10952}}.

\bibitem{Cheng:2019ggx}
H.-Y. Cheng and C.-W. Chiang,
  \ifthenelse{\boolean{articletitles}}{\emph{{Revisiting \CP violation in $D\to
  P\!P$ and $V\!P$ decays}},
  }{}\href{https://doi.org/10.1103/PhysRevD.100.093002}{Phys.\ Rev.\ D
  \textbf{100} (2019) 093002},
  \href{http://arxiv.org/abs/1909.03063}{{\normalfont\ttfamily
  arXiv:1909.03063}}.

\bibitem{Buras:2021rdg}
A.~J. Buras, P.~Colangelo, F.~De~Fazio, and F.~Loparco,
  \ifthenelse{\boolean{articletitles}}{\emph{{The charm of 331}},
  }{}\href{https://doi.org/10.1007/JHEP10(2021)021}{JHEP \textbf{10} (2021)
  021}, \href{http://arxiv.org/abs/2107.10866}{{\normalfont\ttfamily
  arXiv:2107.10866}}.

\bibitem{Schacht:2021jaz}
S.~Schacht and A.~Soni, \ifthenelse{\boolean{articletitles}}{\emph{{Enhancement
  of charm \CP violation due to nearby resonances}},
  }{}\href{https://doi.org/10.1016/j.physletb.2021.136855}{Phys.\ Lett.\ B
  \textbf{825} (2022) 136855},
  \href{http://arxiv.org/abs/2110.07619}{{\normalfont\ttfamily
  arXiv:2110.07619}}.

\bibitem{Bediaga:2022sxw}
I.~Bediaga, T.~Frederico, and P.~C. Magalh\~aes,
  \ifthenelse{\boolean{articletitles}}{\emph{{Enhanced Charm \CP Asymmetries
  from Final State Interactions}},
  }{}\href{https://doi.org/10.1103/PhysRevLett.131.051802}{Phys.\ Rev.\ Lett.\
  \textbf{131} (2023) 051802},
  \href{http://arxiv.org/abs/2203.04056}{{\normalfont\ttfamily
  arXiv:2203.04056}}.

\bibitem{Pich:2023kim}
A.~Pich, E.~Solomonidi, and L.~Vale~Silva,
  \ifthenelse{\boolean{articletitles}}{\emph{{Final-state interactions in the
  CP asymmetries of charm-meson two-body decays}},
  }{}\href{https://doi.org/10.1103/PhysRevD.108.036026}{Phys.\ Rev.\ D
  \textbf{108} (2023) 036026},
  \href{http://arxiv.org/abs/2305.11951}{{\normalfont\ttfamily
  arXiv:2305.11951}}.

\bibitem{Gavrilova:2023fzy}
M.~Gavrilova, Y.~Grossman, and S.~Schacht,
  \ifthenelse{\boolean{articletitles}}{\emph{{Determination of the $D\to\pi\pi$
  ratio of penguin over tree diagrams}},
  }{}\href{https://doi.org/10.1103/PhysRevD.109.033011}{Phys.\ Rev.\ D
  \textbf{109} (2024) 033011},
  \href{http://arxiv.org/abs/2312.10140}{{\normalfont\ttfamily
  arXiv:2312.10140}}.

\bibitem{Lenz:2023rlq}
A.~Lenz, M.~L. Piscopo, and A.~V. Rusov,
  \ifthenelse{\boolean{articletitles}}{\emph{{Two body non-leptonic $D^{0}$
  decays from LCSR and implications for ${\Delta
  a}_{{\text{CP}}}^{{\text{dir}}}$}},
  }{}\href{https://doi.org/10.1007/JHEP03(2024)151}{JHEP \textbf{03} (2024)
  151}, \href{http://arxiv.org/abs/2312.13245}{{\normalfont\ttfamily
  arXiv:2312.13245}}.

\bibitem{Schacht:2022}
S.~Schacht, \ifthenelse{\boolean{articletitles}}{\emph{{A U-spin anomaly in
  charm CP violation}}, }{}\href{https://doi.org/10.1007/JHEP03(2023)205}{JHEP
  \textbf{03} (2023) 205},
  \href{http://arxiv.org/abs/2207.08539}{{\normalfont\ttfamily
  arXiv:2207.08539}}.

\bibitem{Grossman:2012}
Y.~Grossman, A.~L. Kagan, and J.~Zupan,
  \ifthenelse{\boolean{articletitles}}{\emph{{Testing for new physics in singly
  Cabibbo suppressed D decays}},
  }{}\href{https://doi.org/10.1103/PhysRevD.85.114036}{Phys.\ Rev.\ D
  \textbf{85} (2012) 114036},
  \href{http://arxiv.org/abs/1204.3557}{{\normalfont\ttfamily
  arXiv:1204.3557}}.

\bibitem{Bevan:2013xla}
A.~J. Bevan and B.~T. Meadows,
  \ifthenelse{\boolean{articletitles}}{\emph{{Bounding hadronic uncertainties
  in $c\to u$ decays}},
  }{}\href{https://doi.org/10.1103/PhysRevD.90.094028}{Phys.\ Rev.\ D
  \textbf{90} (2014) 094028},
  \href{http://arxiv.org/abs/1310.0050}{{\normalfont\ttfamily
  arXiv:1310.0050}}.

\bibitem{Wang:2022}
D.~Wang, \ifthenelse{\boolean{articletitles}}{\emph{{Evidence of $A_{CP}(D^0\to
  \pi^+\pi^-)$ implies observable $CP$ violation in the $D^0\to \pi^0\pi^0$
  decay}}, }{}\href{https://doi.org/10.1140/epjc/s10052-023-11439-5}{Eur.\
  Phys.\ J.\ C \textbf{83} (2023) 279},
  \href{http://arxiv.org/abs/2207.11053}{{\normalfont\ttfamily
  arXiv:2207.11053}}.

\bibitem{HFLAV}
HFLAV group, Y.~S. Amhis {\em et~al.},
  \ifthenelse{\boolean{articletitles}}{\emph{{Averages of $b$-hadron,
  $c$-hadron, and $\tau$-lepton properties as of 2021}},
  }{}\href{https://doi.org/10.1103/PhysRevD.107.052008}{Phys.\ Rev.\ D
  \textbf{107} (2023) 052008},
  \href{http://arxiv.org/abs/2206.07501}{{\normalfont\ttfamily
  arXiv:2206.07501}}, {updated results and plots available at
  \url{https://hflav.web.cern.ch/}}.

\bibitem{Nisar:2014}
Belle collaboration, N.~K. Nisar {\em et~al.},
  \ifthenelse{\boolean{articletitles}}{\emph{{Search for \CP violation in $D^0
  \to \pi^0 \pi^0$ decays}},
  }{}\href{https://doi.org/10.1103/PhysRevLett.112.211601}{Phys.\ Rev.\ Lett.\
  \textbf{112} (2014) 211601},
  \href{http://arxiv.org/abs/1404.1266}{{\normalfont\ttfamily
  arXiv:1404.1266}}.

\bibitem{LHCb:2021vmn}
LHCb collaboration, R.~Aaij {\em et~al.},
  \ifthenelse{\boolean{articletitles}}{\emph{{Search for time-dependent \CP
  violation in $D^0 \to K^+ K^-$ and $D^0 \to \pi^+ \pi^-$ decays}},
  }{}\href{https://doi.org/10.1103/PhysRevD.104.072010}{Phys.\ Rev.\ D
  \textbf{104} (2021) 072010},
  \href{http://arxiv.org/abs/2105.09889}{{\normalfont\ttfamily
  arXiv:2105.09889}}.

\bibitem{CDF:2011ejf}
CDF collaboration, T.~Aaltonen {\em et~al.},
  \ifthenelse{\boolean{articletitles}}{\emph{{Measurement of \CP-violating
  asymmetries in $D^0\to\pi^+\pi^-$ and $D^0\to K^+K^-$ decays at CDF}},
  }{}\href{https://doi.org/10.1103/PhysRevD.85.012009}{Phys.\ Rev.\ D
  \textbf{85} (2012) 012009},
  \href{http://arxiv.org/abs/1111.5023}{{\normalfont\ttfamily
  arXiv:1111.5023}}.

\bibitem{Berends:1973fd}
F.~A. Berends, K.~J.~F. Gaemers, and R.~Gastmans,
  \ifthenelse{\boolean{articletitles}}{\emph{{$\alpha^3$ contribution to the
  angular asymmetry in $e^+e^-\to\mu^+\mu^-$}},
  }{}\href{https://doi.org/10.1016/0550-3213(73)90153-3}{Nucl.\ Phys.\ B
  \textbf{63} (1973) 381}.

\bibitem{Brown:1973ji}
R.~W. Brown, K.~O. Mikaelian, V.~K. Cung, and E.~A. Paschos,
  \ifthenelse{\boolean{articletitles}}{\emph{{Electromagnetic background in the
  search for neutral weak currents via $e^+e^-\to\mu^+\mu^-$}},
  }{}\href{https://doi.org/10.1016/0370-2693(73)90384-5}{Phys.\ Lett.\ B
  \textbf{43} (1973) 403}.

\bibitem{Cashmore:1985vp}
R.~J. Cashmore, C.~M. Hawkes, B.~W. Lynn, and R.~G. Stuart,
  \ifthenelse{\boolean{articletitles}}{\emph{{The forward-backward asymmetry in
  $e^+ e^- \to \mu^+ \mu^-$ comparisons between the theoretical calculations at
  the one loop level in the Standard Model and with the experimental
  measurements}}, }{}\href{https://doi.org/10.1007/BF01560685}{Z.\ Phys.\ C
  \textbf{30} (1986) 125}.

\bibitem{b2tech}
Belle II collaboration, T.~Abe {\em et~al.},
  \ifthenelse{\boolean{articletitles}}{\emph{{Belle II Technical Design
  Report}}, }{}\href{http://arxiv.org/abs/1011.0352}{{\normalfont\ttfamily
  arXiv:1011.0352}}.

\bibitem{Kou:2018nap}
W.~Altmannshofer {\em et~al.}, \ifthenelse{\boolean{articletitles}}{\emph{{The
  Belle~II physics book}}, }{}\href{https://doi.org/10.1093/ptep/ptz106}{PTEP
  \textbf{2019} (2019) 123C01}, Erratum
  \href{https://doi.org/10.1093/ptep/ptaa008}{ibid.\   \textbf{2020} (2020)
  029201}, \href{http://arxiv.org/abs/1808.10567}{{\normalfont\ttfamily
  arXiv:1808.10567}}.

\bibitem{Akai:2018mbz}
K.~Akai, K.~Furukawa, and H.~Koiso,
  \ifthenelse{\boolean{articletitles}}{\emph{{SuperKEKB collider}},
  }{}\href{https://doi.org/10.1016/j.nima.2018.08.017}{Nucl.\ Instrum.\ Meth.\
  A \textbf{907} (2018) 188},
  \href{http://arxiv.org/abs/1809.01958}{{\normalfont\ttfamily
  arXiv:1809.01958}}.

\bibitem{Lange:2001uf}
D.~J. Lange, \ifthenelse{\boolean{articletitles}}{\emph{{The EvtGen particle
  decay simulation package}},
  }{}\href{https://doi.org/10.1016/S0168-9002(01)00089-4}{Nucl.\ Instrum.\
  Meth.\ A \textbf{462} (2001) 152}.

\bibitem{Jadach:1999vf}
S.~Jadach, B.~F.~L. Ward, and Z.~W\c{a}s,
  \ifthenelse{\boolean{articletitles}}{\emph{{The precision Monte Carlo event
  generator KK for two-fermion final states in $e^+e^-$ collisions}},
  }{}\href{https://doi.org/10.1016/S0010-4655(00)00048-5}{Comput.\ Phys.\
  Commun.\  \textbf{130} (2000) 260},
  \href{http://arxiv.org/abs/hep-ph/9912214}{{\normalfont\ttfamily
  arXiv:hep-ph/9912214}}.

\bibitem{Sjostrand:2014zea}
T.~Sj\"{o}strand {\em et~al.}, \ifthenelse{\boolean{articletitles}}{\emph{{An
  Introduction to PYTHIA 8.2}},
  }{}\href{https://doi.org/10.1016/j.cpc.2015.01.024}{Comput.\ Phys.\ Commun.\
  \textbf{191} (2015) 159},
  \href{http://arxiv.org/abs/1410.3012}{{\normalfont\ttfamily
  arXiv:1410.3012}}.

\bibitem{Barberio:1990ms}
E.~Barberio, B.~van Eijk, and Z.~W\c{a}s,
  \ifthenelse{\boolean{articletitles}}{\emph{{PHOTOS: A universal Monte Carlo
  for QED radiative corrections in decays}},
  }{}\href{https://doi.org/10.1016/0010-4655(91)90012-A}{Comput.\ Phys.\
  Commun.\  \textbf{66} (1991) 115}.

\bibitem{Barberio:1993qi}
E.~Barberio and Z.~W\c{a}s, \ifthenelse{\boolean{articletitles}}{\emph{{PHOTOS:
  A Universal Monte Carlo for QED radiative corrections. Version 2.0}},
  }{}\href{https://doi.org/10.1016/0010-4655(94)90074-4}{Comput.\ Phys.\
  Commun.\  \textbf{79} (1994) 291}.

\bibitem{Agostinelli:2002hh}
GEANT4 collaboration, S.~Agostinelli {\em et~al.},
  \ifthenelse{\boolean{articletitles}}{\emph{{GEANT4: A simulation toolkit}},
  }{}\href{https://doi.org/10.1016/S0168-9002(03)01368-8}{Nucl.\ Instrum.\
  Meth.\  \textbf{A506} (2003) 250}.

\bibitem{Kuhr:2018lps}
Belle II Framework Software Group, T.~Kuhr {\em et~al.},
  \ifthenelse{\boolean{articletitles}}{\emph{{The Belle II Core Software}},
  }{}\href{https://doi.org/10.1007/s41781-018-0017-9}{Comput.\ Softw.\ Big
  Sci.\  \textbf{3} (2019) 1},
  \href{http://arxiv.org/abs/1809.04299}{{\normalfont\ttfamily
  arXiv:1809.04299}}.

\bibitem{basf2-zenodo}
{Belle II collaboration}, \ifthenelse{\boolean{articletitles}}{\emph{{Belle II
  Analysis Software Framework (basf2)}}, }{}
  \url{https://doi.org/10.5281/zenodo.5574115}.

\bibitem{Longo:2020zqt}
S.~Longo {\em et~al.}, \ifthenelse{\boolean{articletitles}}{\emph{{CsI(Tl)
  pulse shape discrimination with the Belle~II electromagnetic calorimeter as a
  novel method to improve particle identification at electron-positron
  colliders}}, }{}\href{https://doi.org/10.1016/j.nima.2020.164562}{Nucl.\
  Instrum.\ Meth.\ A \textbf{982} (2020) 164562},
  \href{http://arxiv.org/abs/2007.09642}{{\normalfont\ttfamily
  arXiv:2007.09642}}.

\bibitem{Krohn:2019dlq}
Belle II Analysis Software Group, J.-F. Krohn {\em et~al.},
  \ifthenelse{\boolean{articletitles}}{\emph{{Global decay chain vertex fitting
  at Belle II}}, }{}\href{https://doi.org/10.1016/j.nima.2020.164269}{Nucl.\
  Instrum.\ Meth.\ A \textbf{976} (2020) 164269},
  \href{http://arxiv.org/abs/1901.11198}{{\normalfont\ttfamily
  arXiv:1901.11198}}.

\bibitem{scikit-learn}
F.~Pedregosa {\em et~al.},
  \ifthenelse{\boolean{articletitles}}{\emph{Scikit-learn: Machine learning in
  {P}ython}, }{}J.\ Mach.\ Learn.\ Res.\  \textbf{12} (2011) 2825,
  \href{http://arxiv.org/abs/1201.0490}{{\normalfont\ttfamily
  arXiv:1201.0490}}.

\bibitem{histogram-gbdt}
A.~Bellet, N.~Morvan, and J.~Salmon,
  \ifthenelse{\boolean{articletitles}}{\emph{Scikit-learn: Machine learning
  without learning the gradient}, }{} in {\em Advances in Neural Information
  Processing Systems}, 2018.

\bibitem{PID}
Belle II collaboration, I.~Adachi {\em et~al.},
  \ifthenelse{\boolean{articletitles}}{\emph{{Charged-hadron identification at
  Belle~II}}, }{}\href{http://arxiv.org/abs/TBD}{{\normalfont\ttfamily
  arXiv:TBD}}.

\bibitem{Rogozhnikov:2016bdp}
A.~Rogozhnikov, \ifthenelse{\boolean{articletitles}}{\emph{{Reweighting with
  Boosted Decision Trees}},
  }{}\href{https://doi.org/10.1088/1742-6596/762/1/012036}{J.\ Phys.\ Conf.\
  Ser.\  \textbf{762} (2016) 012036},
  \href{http://arxiv.org/abs/1608.05806}{{\normalfont\ttfamily
  arXiv:1608.05806}}.

\bibitem{Pivk:2004ty}
M.~Pivk and F.~R. Le~Diberder,
  \ifthenelse{\boolean{articletitles}}{\emph{{sPlot: A statistical tool to
  unfold data distributions}},
  }{}\href{https://doi.org/10.1016/j.nima.2005.08.106}{Nucl.\ Instrum.\ Meth.\
  A \textbf{555} (2005) 356},
  \href{http://arxiv.org/abs/physics/0402083}{{\normalfont\ttfamily
  arXiv:physics/0402083}}.

\bibitem{johnson}
N.~L. Johnson, \ifthenelse{\boolean{articletitles}}{\emph{{Systems of frequency
  curves generated by methods of translation}},
  }{}\href{https://doi.org/10.1093/biomet/36.1-2.149}{Biometrika \textbf{36}
  (1949) 149}.

\bibitem{pdg}
Particle Data Group, S.~Navas {\em et~al.},
  \ifthenelse{\boolean{articletitles}}{\emph{{Review of particle physics}},
  }{}\href{https://doi.org/10.1103/PhysRevD.110.030001}{Phys.\ Rev.\ D
  \textbf{110} (2024) 030001}.

\end{thebibliography}
\providecommand{\href}[2]{#2}\begingroup\raggedright\endgroup

\end{document}